 \documentclass[aps,prd,amsfonts,floats,floatfix,eqsecnum,notitlepage,nofootinbib,superscriptaddress]{revtex4-2}
\usepackage[dvipsnames]{xcolor}
\usepackage{natbib}
\usepackage{amsmath}
\usepackage{amssymb}

\usepackage[normalem]{ulem}
\usepackage[colorlinks=true, citecolor=blue,urlcolor=blue]{hyperref}

\usepackage[retainorgcmds]{IEEEtrantools}
\usepackage{graphicx}
\usepackage{graphics}
\usepackage{color}
\usepackage{subfig}

\usepackage{ulem}
\graphicspath{{../Plots/}}

\newtheorem{proposition}{Proposition}

\newcommand{\eqn}[1]
{
    \begin{equation}#1\end{equation}
}

\newcommand{\eqnalgn}[1]
{
    \begin{align}#1\end{align}
}
\newcommand{\Dt}[0]{\ensuremath{\Delta t}}
\newcommand{\lC}{\left[}
\newcommand{\rC}{\right]}
\newcommand{\lP}{\left(}
\newcommand{\rP}{\right)}

\newcommand{\df}{\textrm{d}}

\newcommand{\dd}[2]{\frac{\textrm{d}#1}{\textrm{d}#2}}

\newcommand{\TOp}{{}_s\mathcal{T}}

\newcommand{\s}{{}_{s}}
\newcommand{\dt}{\Delta t}

\newcommand{\Gret}[1]{{}_{#1}G_{\text{ret}}}
\newcommand{\Gd}[1]{{}_{#1}G_{\text{d}}}
\newcommand{\Gnd}[1]{{}_{#1}G_{\text{nd}}}
\newcommand{\sYlm}[0]{\ensuremath{{}_{s}Y_{\ell m}}}

\newcommand{\GretT}{_{s}G_{\textrm{ret}}}

\newcommand{\indmode}{\ell}

\newcommand{\sGTld}[1]{{}_{#1}G^\text{d}_{\indmode}}
\newcommand{\sGTldt}[1]{{}_{#1}\tilde{G}^\text{d}_{\indmode}}
\newcommand{\sGTl}[1]{{}_{#1}G_{\indmode}}

\newcommand{\be}{\begin{equation}}
\newcommand{\ee}{\end{equation}}

\newcommand{\cd}{\mathcal{D}}

\newcommand{\real}{\mathbb{R}}
\newcommand{\ep}{E}

\begin{document}
\global\parskip 6pt

\author{David Q. Aruquipa}
\email{david.q.aruquipa@gmail.com}
\noaffiliation

\author{Marc Casals}
\email{marc.casals@uni-leipzig.de}
\affiliation{
Institut f\"ur Theoretische Physik, Universit\"at Leipzig,\\ Br\"uderstra{\ss}e 16, 04103 Leipzig, Germany}
\affiliation{Scoil na Matamaitice agus na Staitistic\'i, An Col\'aiste Ollscoile, Baile \'Atha Cliath,  D04 V1W8, Ireland}
\affiliation{Centro Brasileiro de Pesquisas F\'isicas (CBPF), Rio de Janeiro, CEP 22290-180, Brazil}

\author{Brien C. Nolan}
\email{brien.nolan@dcu.ie}
\affiliation{School of Mathematical Sciences, Dublin City
University, Glasnevin, Dublin 9, Ireland.}

\title{Calculation of a regularized   Teukolsky Green function in Schwarzschild spacetime}

\begin{abstract}
We obtain exact expressions for various factors involved in
the Hadamard form of the retarded Green function for the (Bardeen-Press-)Teukolsky equation on Schwarzschild spacetime.  We use these  to improve  on previous results for the calculation of this  Green function.
We work in a spacetime $\mathcal{M}_2\times\mathbb{S}^2$ conformal to Schwarzschild, in which the metric takes a direct product form. This allows us to derive a separable form for the direct (i.e., singular) part of the Hadamard form of the retarded Green function. The angular factor in this quantity is calculated explicitly. This shows an interesting interplay between geodesics of $\mathbb{S}^2$, spin-weighted spherical harmonics, and Euler angles. The $\mathcal{M}_2$ factor equates to a spin-dependent factor that satisfies a transport equation along geodesics, times the square root of the van Vleck determinant. Both terms are calculated in an exact form  for constant radius orbits (which includes the cases  of circular timelike geodesics and static worldlines of Schwarzschild spacetime).
This separable form also allows us to obtain the multipolar $\ell$-modes of the direct part for electromagnetic and gravitational field perturbations. We then use these $\ell$-modes to calculate, in the gravitational case, the retarded Green function minus its direct part: this is a better representation in practise of the retarded Green function for points near coincidence.
\end{abstract}

\maketitle


\section{Introduction}

Performing practical calculations of the $4$-dimensional retarded Green function (GF) for field perturbations of black hole spacetimes is notoriously difficult. One of the main technical difficulties is that the GF is a (two-point) distribution on spacetime, with singular support at coincidence and along any null geodesic connecting the two points \cite{Duistermaat1972Propagation}. It is often useful to calculate the GF by decomposing it into multipolar $\ell$-modes, which obey a differential equation of dimension lower than that obeyed by the full GF. This can be extremely effective in some regimes. However, as this approach {necessarily} involves {in practice} a {\it finite} mode-sum, it can only provide an approximation that is qualitatively different to the GF on its singular support: for example, Dirac$-\delta$ distributions {become} smeared to Gaussian peaks. This smearing `contaminates' the calculation away from the singularities of the GF {calculated} via a multipolar decomposition.
As an example of a consequence of that, in the scalar field case, this contamination can hinder the calculation of the scalar self-field and self-force, when obtained as worldline integrals of, respectively, the GF and its derivatives (see, e.g.,~\cite{poisson2011motion} for a review of the self-force  and~\cite{CDOW13,PhysRevD.89.084021} for the calculation of the scalar self-field and self-force via worldline integrals and using a multipolar decomposition of the GF).

Fortunately, the explicit form of the  GF (including its singular part) near coincidence is provided by the \textit{Hadamard form}~\cite{Hadamard,friedlander1975wave}: the singular part is given by a  Dirac-$\delta$ distribution with  support at coincidence and when the points are lightlike-separated. The Hadamard form thus allows for finding alternative methods for calculating the GF near coincidence.
One of these methods is to calculate the multipolar modes of the divergent Dirac-$\delta$ term (called the direct term) in the Hadamard form and subtract them from the modes of the full GF.
We refer to the summation of the modes resulting from this subtraction as the non-direct part of the GF.
Since the Dirac-$\delta$ term  has sharp support where its argument is zero,  the non-direct part of the GF
should  provide a better approximation to the GF at any pairs of points except for exactly at coincidence (or for points connected by a direct null geodesic). The singular part of the GF  exactly at coincidence is not needed for, e.g., self-field/force calculations since it is removed as part of a regularization procedure. If the singular part of the GF at coincidence is needed (via, e.g., an integral of the GF) for other applications, then this can be calculated separately (using, e.g., the Hadamard form again) and surgically added separately (this  was done, e.g., for  communication between quantum particle detectors in~\cite{jonsson2020communication}). 

The described method for improving the calculation of the GF away from coincidence was suggested and implemented in Ref.~\cite{CNOW2019} in the case of scalar field perturbations of Schwarzschild spacetime. This was achieved by leveraging the fact that Schwarzschild spacetime is conformal to the direct product of a $2$-dimensional spacetime ($\mathcal{M}_2$) and the $2$-dimensional sphere $\mathbb{S}^2$. In the scalar case, this property was exploited in~\cite{casals2023global} for finding a  representation for the GF which makes explicit its complete  singularity structure and then used in~\cite{CNOW2019} for obtained the multipolar modes of the direct part in the Hadamard form  and then calculating the non-direct part of the (scalar) GF.

In this paper we carry out a similar calculation for the case of {\it generic}-spin field perturbations of Schwarzschild space-time, which satisfy the (Bardeen-Press-)Teukolsky (BPT) equation~\cite{barden-doi:10.1063/1.1666175,teukolsky1972rotating,Teukolsky:1973ha}. 
Specifically, we factor out the direct Hadamard term  into two factors, one coming from $\mathcal{M}_2$ and the other from  $\mathbb{S}^2$. The factor coming from the $\mathbb{S}^2$ we obtain in closed form in terms of Euler angles (see Eq.~\eqref{uring-sol}). As for the factor coming from the 
$\mathcal{M}_2$, we express it in terms of its value in the scalar case (which is the square root of the  van Vleck determinant~\cite{poisson2011motion,casals2023global} in $\mathcal{M}_2$) times a coefficient which is written as an integral along geodesics in $\mathcal{M}_2$ (see Eq.~\eqref{lambda-final}).
We thus provide a fully geometrical expression for the direct Hadamard term for the BPT equation  lifted to the conformal Schwarzschild spacetime. We then use this representation in order to obtain an analytic expression for the multipolar modes of the direct Hadamard term. We calculate these modes in the cases of electromagnetic (spin $s=-1$) and gravitational ($s=-2$) perturbations. We finally use this calculation together with a calculation provided in~\cite{aruquipa2026greenfunctionsreggewheelerteukolsky} of the multipolar modes of the full GF for $s=-2$ to calculate the non-direct part of this gravitational GF.
In certain restrictive cases, we calculate the direct term in the Hadamard form directly in Schwarzschild spacetime (without resorting to the mentioned $2+2$ decomposition) either fully in closed form or else in terms of the van Vleck determinant in Schwarzschild; the above $2+2$ decomposition is instead useful for generic spacetime points and for calculating the $\ell$-modes (with the consequent improvement in the representation of the GF away from coincidence).

The rest of this paper consists of the following. In Sec.~\ref{sec:Had} we present the Hadamard form for the GF in Schwarzschild spacetime, provide the equations satisfied by its various biscalars and obtain analytic expressions for its direct term along specific worldlines. We start  using the $2+2$ approach in Sec.~\ref{sec:2+2 BPT}, where we express the BPT equation in its $2+2$ form. In Sec.~\ref{sec:Had-confSchw} we derive analytic expressions for the $\mathcal{M}_2$-  and $\mathbb{S}^2$-factors in the direct part of the Hadamard form. We use these  expressions for the factors in Sec.~\ref{sec:non-direct}
 in order to obtain an analytic expression for the $\ell$-modes of the direct part, calculate these modes explicitly for spin $s=-1$ and $-2$ and then use the latter to calculate the non-direct part of the BPT GF for $s=-2$  in Schwarzschild spacetime. 
 Finally, in appendices \ref{sec:AppA}--\ref{sec:SmallCoordExpansionForBarU} we provide representations and expansions for Hadamard quantities in  $\mathcal{M}_2$; in App.~\ref{app:sVAtCoincidence} we calculate the coincidence limit of the non-direct (tail) Hadamard coefficient in Schwarzschild.

Throughout this paper we use geometric units $c= G=1$, and we use the conventions of Ref.~\cite{wald1984general}.

\section{Hadamard form of the BPT retarded Green function}\label{sec:Had}

The exterior Schwarzschild spacetime $(M,g)$ is given by $M=\real\times(2M,+\infty)\times\mathbb{S}^2$, and the line element, in Schwarzschild coordinates $(x^\mu)=(t,r,\theta,\phi)$,
\be ds^2=g_{\mu\nu}dx^\mu dx^\nu = -fdt^2+f^{-1}dr^2+r^2d\Omega^2_2,\label{metric} \ee
where $f=f(r)=1-\frac{2M}{r}$ and  $d\Omega_2^2=d\theta^2+\sin^2\theta d\phi^2$ is the line element of $\mathbb{S}^2$.

Helicity$-s$   fields ($s=0$, $\pm 1$ and $\pm 2$ corresponding to scalar, electromagnetic and gravitational perturbations, respectively) on Schwarzschild spacetime satisfy the  celebrated (Bardeen-Press-)Teukolsky (BPT) equation~\cite{barden-doi:10.1063/1.1666175,teukolsky1972rotating,Teukolsky:1973ha}.
The spin of the field is $|s|$, but following common usage, we shall refer to $s$ as the spin instead of helicity.
The retarded Green function (GF) $\GretT$ of the BPT equation satisfies
\be _{s}\mathcal{T}( \GretT(x,x'))=-4\pi\delta_4(x,x'),\label{GT-ret} \ee
where the Teukolsky operator is 
\be _s\mathcal{T}=
\Box+A^\mu\partial_\mu+B=
\Box+\frac{2s}{r^2}(a^\mu\partial_\mu+b), \label{sT-op} \ee
and $\delta_4(x,x')$ is the Dirac delta function on $M$.
Here, $\Box$ is the d'Alembertian of the Schwarzschild metric tensor $g$, and we have defined
\eqn{\label{eqn:ABEqns}
A^{\mu}=\frac{2s}{r^2}a^{\mu},\quad B=\frac{2s}{r^2}b,
}
with
\begin{eqnarray} 
a^\mu\partial_\mu &=& -\frac{r}{f}\left(1-\frac{3M}{r}\right)\partial_t+(r-M)\partial_r+i\cot\theta\csc\theta\partial_\phi,
\label{A-vec}\\
b &=& \frac12(1-s\cot^2\theta). \label{b-scalar}
\end{eqnarray}
The ``retarded" boundary condition on $_{s}G_{\textrm{ret}}(x,x')$ requires that it vanishes when the field point $x$ does not lie to the causal future of the base point $x'$.

In local, normal neighbourhoods the so-called Hadamard form for the GF $\GretT(x,x')$ provides a representation which makes explicit its singular nature when the points $x$ and $x'$ are connected via a null geodesic.
A normal neighbourhood $\mathcal{N}(x)$ of a point $x$ is a neighbourhood of $x$ such that
every $x'\in \mathcal{N}(x)$ is connected to $x$ by a unique geodesic which
lies in $\mathcal{N}(x)$.
Within a normal neighbourhood of $x'$, the Hadamard form for $\GretT(x,x')$ is  (see e.g.\ Theorem 4.5.1 of \cite{friedlander1975wave}) 
\eqn{\label{eqn:hadamardTGret}
   \GretT(x,x')={}_sU(x,x')\delta_+(\sigma)+{}_sV(x,x')\theta_+(-\sigma),
}
where ${}_sU(x,x')$ and ${}_sV(x,x')$ are  smooth biscalars, 
$\sigma=\sigma(x,x')$ is  Synge's world function 
(i.e., it is equal to half of the (signed) squared distance along the unique geodesic connecting $x$ and $x'$). 
 $\delta_+$ and $\theta_+$ are the spacetime delta- and step-functions, respectively, supported in the region $\{t>t'\}$.
 The terms ${}_sU\delta_+$ and ${}_sV\theta_+$  in \eqref{eqn:hadamardTGret}  are referred to as, respectively, the direct and tail Hadamard terms.

Following Ref.~\cite{friedlander1975wave} (see also~\cite{aruquipa2026greenfunctionsreggewheelerteukolsky,2023PhRvD.108l5017I}) the direct Hadamard biscalar ${}_sU$ is taken to satisfy the transport equation 
\eqn{\label{eqn:transportUT}
    \lP2\sigma^\mu\partial_\mu+{\sigma^\mu}_\mu+\sigma_\mu A^\mu -4\rP {}_sU=0.}
   As usual \cite{poisson2011motion}, we use $\sigma_\mu=\nabla_\mu\sigma$ and ${\sigma^\mu}_\mu=\Box\sigma$. Also, the tail Hadamard biscalar ${}_sV$ satisfies the homogeneous Teukolsky equation,
 \eqn{\label{eqn:VTs}
    \TOp\,
    ({}_sV)=0,}
complemented by the following  boundary condition as a transport equation (this agrees with the corresponding equation of \cite{2023PhRvD.108l5017I}, but note that ${}_sV$ in this paper has the opposite sign to the corresponding term in the present paper):
\begin{align}\label{eqn:VTBoundaryConditions}
    2\sigma^\mu\partial_\mu\left({}_s\check{V}\right)+({\sigma^\mu}_\mu+\sigma_\mu A^\mu-2){}_s\check{V}=
\left.
    \lP
 \TOp\,
    ({}_sU)\rP\right|_{\sigma=0},
\end{align}
where ${}_s\check{V}\equiv \left.{}_sV\right|_{\sigma=0}$.
In their turn, the transport equations \eqref{eqn:transportUT} for ${}_sU$ and  \eqref{eqn:VTBoundaryConditions} for ${}_sV$ are complemented by the initial conditions, respectively,
\eqn{\label{eqn:UTInitialCondition}
    \lC {}_sU
    \rC=
    1
}
and
\eqn{\label{eqn:hatVTsInitialCondition}
    \lC{}_s\check{V}\rC=\lC {}_sV\rC=\frac{1}{2}\lC\Box {}_sU\rC+\frac{1}{2} \lC A^\mu\partial_{\mu}\lP{}_sU\rP\rC+\frac{1}{2} B.
} 
We use square brackets around a bitensor $A(x,x')$ to denote its value  at coincidence (i.e., $[A]=A(x,x)$).

In the scalar case $s=0$, the direct biscalar is simply equal to the square root of the van Vleck determinant $\Delta$ on Schwarzschild spacetime:
\eqn{\label{eq:U=VV^1/2}
{}_0U(x,x')=\Delta^{1/2}(x,x').
}

We note that, whereas for $s=0$, 
${}_sU(x,x')$, ${}_sV(x,x')$ and $\GretT(x,x')$ are both symmetric and real-valued and ${}_s\mathcal{T}$ is a self-adjoint operator, none of these properties need to hold for $s\neq 0$. 

In App.~\ref{app:sVAtCoincidence} we use Eqs.~\eqref{eqn:hatVTsInitialCondition} and~\eqref{eqn:transportUT} to derive a covariant expansion (i.e., in terms of powers of $\sigma^{\mu}$) for ${}_sU$ and to obtain the coincidence limit  
\begin{equation}\label{eq:Vs-coinc}
[{}_sV]=-s^2\frac{2M}{r^3}.
\end{equation}

\subsection{Some exact solutions for the direct term in the Hadamard form}\label{sec:Exact 4-D}

In this subsection, we obtain some analytical, closed form expressions for the direct term ${}_sU$ in the Hadamard form \eqref{eqn:hadamardTGret} for the BPT GF $\GretT$ along specific worldlines. 

With the help of Eqs.~\eqref{eqn:transportUT} and \eqref{eqn:UTInitialCondition} for $s=0$ as well as \eqref{eq:U=VV^1/2}, the transport
 equation~\eqref{eqn:transportUT} for generic-$s$ can be rewritten as
\eqn{\label{eq:Transp-UT}
\lambda \dd{}{\lambda}\lP\ln{\lP{{}_sU}\rP^2}-\ln\Delta\rP+\lambda\,\dot{z}^\mu A_\mu=0,
}
where $\dot{z}^\mu\equiv \dd{z^\mu}{\lambda}$ and $\lambda$ is an affine parameter along a geodesic $z(\lambda)$.
Integrating Eq.~\eqref{eq:Transp-UT}, we obtain
\begin{eqnarray}
    2\ln\lP \frac{{}_sU(x,x')}{\Delta^{1/2}(x,x')}\rP&=&-\int_{\lambda_1}^{\lambda_2}\dot{z^\mu}A_\mu(z(\lambda)) \df\lambda\nonumber \\
    &=&\int_{\lambda_1}^{\lambda_2}\lP f\dot{t}A^t-\dot{r}\frac{A^r}{f}- r^2\dot{\phi}\sin^2\theta A^\phi\rP\df\lambda,\label{eq:lnUT}
\end{eqnarray}
where $x=z(\lambda_2)$ and $x'=z(\lambda_1)$, for some $\lambda_2\geq \lambda_1$.
We may solve the integral in Eq.~\eqref{eq:lnUT} for specific geodesics.

First, we set $\theta=\theta'=\pi/2$, which implies $A^\phi=0$. In some other contexts in Schwarzschild spacetime, this does not imply any loss of generality. For example, any one individual geodesic can always be assumed to be confined to the equatorial plane $\theta=\pi/2$. However, the definition of spin-weighting breaks spherical symmetry, as an axis of the 2-sphere must be chosen (see e.g.\ \cite{goldberg1967spin}).  Hence the assumption that $\theta=\theta'=\pi/2$ is a strong restriction on the points where the BPT GF is evaluated.

We now look at two cases.
The first case is that of a circular geodesic, where $r=r'$ is constant. This yields
\eqn{
2\ln\lP \frac{{}_sU}{\Delta^{1/2}}\rP=fA^t\int_{t'}^{t}\df \bar{t}=fA^t\Delta t,
}
where
\be \Delta t = t-t'.\label{delta-t-def} \ee
It follows that 
\eqn{\label{eq:Us circ}
{}_sU(x,x')=e^{fA^t\Dt/2}\Delta^{1/2}=\exp{\lC{-\frac{s(r-3M)}{r^2}\Delta t}\rC}\Delta^{1/2}\quad \text{along circular geodesics}.
}
We note that the spin-dependent prefactor in front of the spin-0 direct Hadamard bitensor  $\Delta^{1/2}$ in \eqref{eq:Us circ} increases or decreases exponentially with time depending on the sign of $s(r-3M)$, i.e., depending on  the helicity of the field and whether the field point is inside or outside the photon ring.

The second case is that of  radial null geodesics, where
 $\epsilon\, \df t=\df r_*$, with $\epsilon=+1$ and $-1$ for, respectively, outgoing and ingoing geodesics.
\begin{eqnarray}
 2\ln\lP \frac{{}_sU}{\Delta^{1/2}}\rP
&=&\,\int_{r'}^{r}\lP\epsilon\, A^t(\bar{r})-\frac{A^r(\bar{r})}{f(\bar{r})}\rP\df\bar{r}\nonumber\\
&=&\,\ln\lP\frac{r'}{r}\rP^{s(3\epsilon+1)}+\ln\lP\frac{r'-2M}{r-2M}\rP^{s(1-\epsilon)}. \label{eq:lnUT-RadialNull}
\end{eqnarray}
Furthermore, along radial null geodesics we have $\Delta=1$. This arises by considering the radial limit of results of \cite{hollowood2009refractive} for general null geodesics of Schwarzschild spacetime. Hence, Eq.~\eqref{eq:lnUT-RadialNull} yields 
\eqn{\label{eq:UT,radial null}
{}_sU(x,x')=\lP\frac{r'}{r}\rP^{s(3\epsilon+1)/2}\lP\frac{r'-2M}{r-2M}\rP^{s(1-\epsilon)/2}\quad \text{along radial null geodesics}.
}

Further progress taking this 4-dimensional approach seems difficult, and as we have seen, the results above are limited by the assumption that $\theta=\theta'=\pi/2$. Thus we take an alternative approach that exploits properties of a 2+2 conformal decompostion of Schwarzschild spacetime.

\section{2+2 form of the BPT equation}\label{sec:2+2 BPT}

We work in the conformally rescaled spacetime with line element $d\hat{s}^2=r^{-2}ds^2$. The corresponding metric has the direct sum form 
\be \hat{g}_{\mu\nu} = \left(\begin{array}{c|c} \bar{g}_{ij}(x^k) & 0 \\ \hline 0 & \mathring{g}_{AB}(x^C) 
\end{array} \right), \label{g-hat} \ee
where $x^i=(t,r)$ are coordinates on the Lorentzian 2-space $(\mathcal{M}_2,\bar{g})$ and $x^A=(\theta,\phi)$ are coordinates on the unit round sphere $(\mathbb{S}^2,\mathring{g})$. This $4d$ \textit{conformal Schwarzschild} spacetime {$(\hat{M},\hat{g})$, with $\hat{M}=\mathcal{M}_2\times\mathbb{S}^2$,} has Ricci scalar 
\be \hat{R} = \frac{12M}{r},\label{ricci-hat} \ee
and by properties of conformal transformations (see e.g.\ Appendix D of \cite{wald1984general}), we know that 
\be \Box\psi = \frac{1}{r^3}\left(\hat{\Box}-\frac{2M}{r}\right)(r\psi). \label{box-conf} \ee
It follows that defining $\hat{\psi}=r\psi$, we can write the Teukolsky operator in the form 
\be r^3 (_s\mathcal{T}(\psi)) = {_s\hat{\mathcal{T}}}(\hat{\psi}) \label{conf-t-op-trans} \ee
where 
\be _s\hat{\mathcal{T}} = \hat{\Box}+2s\,a^\mu\partial_\mu-\frac{2s}{r}a^\mu\partial_\mu r-\frac{2M}{r}+2sb \label{conf-t-op} \ee
is the \textit{conformal Teukolsky operator}. 
We now apply a conformal transformation to the retarded Green function and define 
\be _s\hat{G}_{\textrm{ret}} {(x,x')}= r\,r'\, _sG_{\textrm{ret}}{(x,x')}.\label{conf-g-ret} \ee
By (\ref{conf-t-op-trans}), \eqref{GT-ret} and conformal properties of the Dirac measure, we find 
\be _s\hat{\mathcal{T}}( _s\hat{G}_{\textrm{ret}}(x,x')) = -4\pi\hat{\delta}(x,x'). \label{Gret-t-conf} \ee
The new Green function $_s\hat{G}_{\textrm{ret}}$ satisfies the same causal properties as $_s{G}_{\textrm{ret}}$ (namely, it vanishes whenever $x$ does not lie to the causal future of $x'$, this being a conformally invariant property), and so we see that it is the {\it retarded} Green function for the conformal Teukolsky operator. 

The $2+2$ direct product structure of (\ref{g-hat}) leads to the decomposition of the $4$d conformal d'Alembertian operator as
 \be \hat{\Box} = \bar{\Box} + \mathring{\bigtriangleup} \label{dalem-decomp} \ee
 where $\bar{\Box}$ is the d'Alembertian of $\bar{g}$ and $\mathring{\bigtriangleup}$ is the Laplacian of $\mathbb{S}^2$. This in turn underpins the well known separability of the Teukolsky equation: we can write  
 \be _s{\hat{\mathcal{T}}} = {_s{\mathcal{P}}}(t,r) + {_s\mathcal{L}}(\theta,\phi) \label{t-op-decomp}\ee
 where
 \begin{align}
 {_s{\mathcal{P}}}(t,r)&= \bar{\Box}+2sa^i\partial_i-2s\left(1-\frac{M}{r}\right)-\frac{2M}{r},\label{conf-split-m2}
 \\ {_s\mathcal{L}}(\theta,\phi)&= \mathring{\bigtriangleup}+2is\frac{\cos\theta}{\sin^2\theta}\partial_\phi+s(1-s\cot^2\theta).\label{conf-split-s2} 
 \end{align}

Our ultimate  aim is to exploit features of the $2+2$ conformal decomposition above to (i) derive analytic features of the retarded Green function $_s\hat{G}_{\textrm{ret}}$ and (ii) to use these to undertake efficient calculations of related quantities - e.g.\ the first order self-field  of a self-gravitating particle in Schwarzschild spacetime. In the present paper, we obtain an analytic representation of ${}_s\hat{U}$, and apply this to a regularized calculation of the GF. These are enabled on the one hand by geometrical features of $(\mathcal{M}_2,\bar{g})$ - principally, the fact that this spacetime is a causal domain \cite{casals2015geometric} (and so in particular each pair of points is connected by a unique geodesic, and the world function is uniquely and globally defined) - and on the other by connections between geodesics, spin-weighted spherical harmonics and Euler angles on $\mathbb{S}^2$. 

\section{Hadamard form of the retarded Green function in conformal Schwarzschild}\label{sec:Had-confSchw}

In this section, we confine ourselves to normal neighbourhoods. Similarly to \eqref{eqn:hadamardTGret} for the retarded Green function of Schwarzschild, 
in a normal neighbourhood  of $x'=(x^{i'},x^{A'})\in \hat{M}$, the retarded Green function of (\ref{Gret-t-conf}) can be written in the Hadamard form \cite{friedlander1975wave} 
\be _s{\hat{G}}_{\textrm{ret}}(x,x') = {_s{\hat{U}}}(x,x')\delta_+(\hat{\sigma}) + {_s{\hat{V}}}(x,x')\theta_+(-\hat{\sigma}) \label{Gret-Had} \ee
where $_s{\hat{U}}$ and $_s{\hat{V}}$ are smooth two-point functions on $\hat{M}$ with spin-weight  $s$ \cite{teukolsky1972rotating}. Due to the direct sum structure of the metric, the world function $\hat{\sigma}$ of $(\hat{M},\hat{g})$ decomposes as \cite{casals2012kirchhoff} 
\be \hat{\sigma}(x,x') = \bar{\sigma}(x^i,x^{i'}) + \mathring{\sigma}(x^A,x^{A'}) \label{sigma-decomp} \ee
where $\bar{\sigma}$ and $\mathring{\sigma}$ are the corresponding world functions of $(\mathcal{M}_2,\bar{g})$ and $(\mathbb{S}^2,\mathring{g})$ respectively, which satisfy
\begin{align}\label{eq:bar-sigma}
    \nabla_i\bar\sigma\nabla^i\bar\sigma&=2\bar\sigma,\\
    \nabla_A\mathring{\sigma}\nabla^A\mathring{\sigma}&=2\mathring{\sigma}.
\end{align}
We note that both of these world functions are defined {\it globally} on their respective space(time)s.

Following Friedlander's approach in~\cite{friedlander1975wave}, the coefficient $_s{\hat{U}}(x,x')$ of the \textit{direct part} of the retarded Green function is taken to satisfy the transport equation 
\be 2\hat{\sigma}^\alpha\hat{\nabla}_\alpha({_s{\hat{U}}})+\left(\hat{\Box}\hat{\sigma}+2sa^\mu\partial_\mu\hat{\sigma}-4\right){_s{\hat{U}}}=0,
\label{u-transport} \ee
where $2sa^\mu\partial_\mu$ is the vector component of the conformal Teukolsky operator. The following coincidence limit also applies: 
\be [{_s{\hat{U}}}]=1.\label{U-coincidence} \ee
The decomposition (\ref{sigma-decomp}) allows us to separate the variables in (\ref{u-transport}). Writing 
\be {_s{\hat{U}}}{(x,x')} = {_s{\bar{U}}}(x^i,x^{i'}){_s{\mathring{U}}}(x^A,x^{A'}), \label{U-sep} \ee
we find
\begin{eqnarray}
2\bar{\sigma}^i\bar{\nabla}_i({_s{\bar{U}}}) + (\bar{\Box}\bar{\sigma}+2sa^i\partial_i\bar{\sigma}-2){_s{\bar{U}}} &=& 0,\label{ubar-transport} \\
2\mathring{\sigma}^A\mathring{\nabla}_A({_s{\mathring{U}}}) + (\mathring{\bigtriangleup}\mathring{\sigma}+2sa^A\partial_A\mathring{\sigma}-2){_s{\mathring{U}}} &=& 0\label{uring-transport} 
\end{eqnarray}
where an arbitrary separation constant has been set equal to zero. We impose the same coincidence limits on both ${_s{\bar{U}}}$ and ${_s{\mathring{U}}}$:
\be [{_s{\bar{U}}}]=[{_s{\mathring{U}}}]=1.\label{u-bar-ring-coincidence} \ee
We note that ${_s{\bar{U}}}$ is a two-point function on $\mathcal{M}_2$ of spin-weight 0, but ${_s{\mathring{U}}}$ is a two-point function on $\mathbb{S}^2$ of spin-weight $s$.    

\subsection{The direct part on $\mathbb{S}^2$.}

In this section, we derive an explicit geometrical formula for ${_s{\mathring{U}}}$. This arises through an interesting connection between the Teukolsky equation and spin-weighted functions, geodesics and Euler angles on $\mathbb{S}^2$. We work in the usual polar-azimuthal coordinates $(\theta,\phi)$. Then the geodesic distance $\gamma(x^A,x^{A'})$ between two points $P:(\theta,\phi)$ and $P':(\theta',\phi')$ on $\mathbb{S}^2$ satisfies 
\be \cos\gamma = \cos\theta\cos\theta'+\sin\theta\sin\theta'\cos(\phi-\phi'). \label{geo-dist-s2} \ee

To make the connection with Euler angles, we consider the usual embedding of $\mathbb{S}^2$ in the Euclidean space $\real^3$. In Cartesian coordinates $(x,y,z)$ (and the associated right-handed frame), we identify $P$ and $P'$ with the position vectors 
\begin{eqnarray}
\vec{p} &=&(\sin\theta\cos\phi,\sin\theta\sin\phi,\cos\theta),\label{p-in-r3} \\
\vec{p}\,' &=&(\sin\theta'\cos\phi',\sin\theta'\sin\phi',\cos\theta').\label{p'-in-r3} 
\end{eqnarray}

\begin{figure}[h!]
\begin{center}
\includegraphics[width=6.2cm]{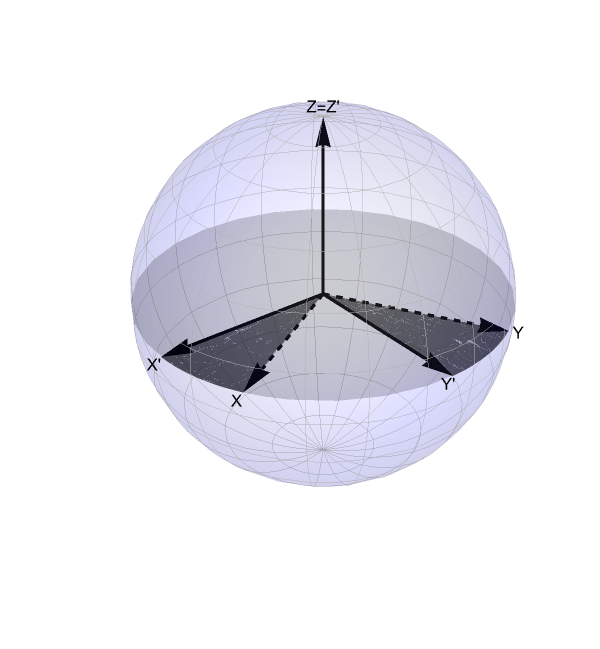}
\hskip-26pt
\includegraphics[width=6.2cm]{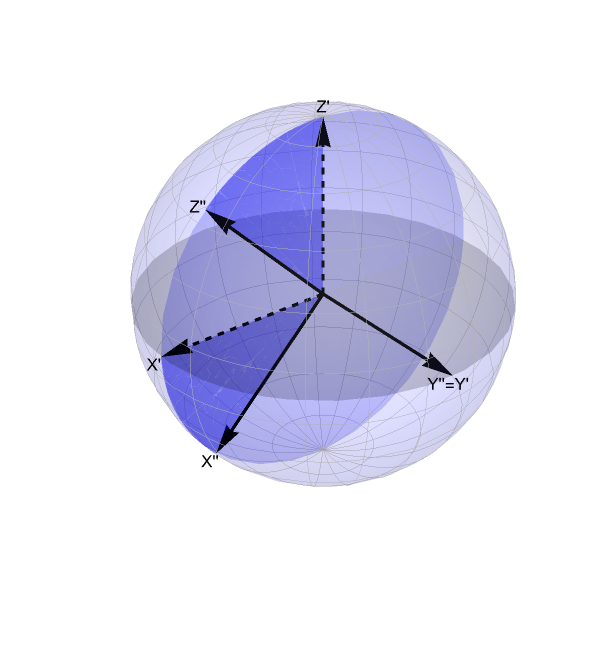}
\hskip-26pt
\includegraphics[width=6.2cm]{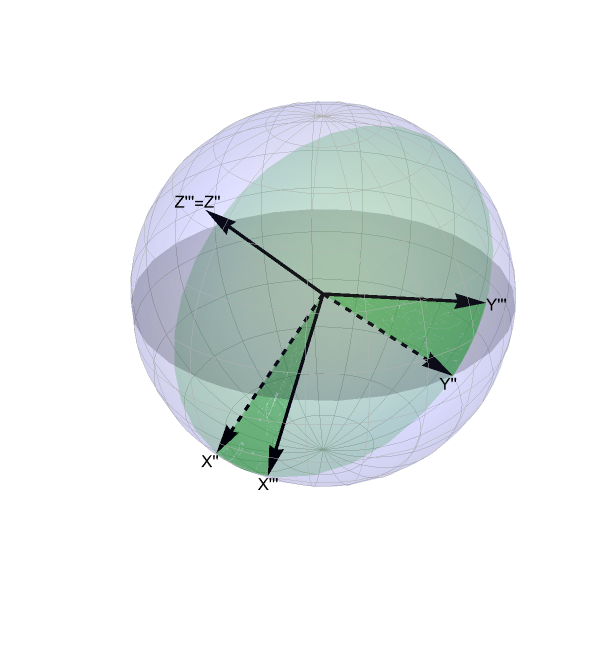}
\end{center}
\caption{The rotations and Euler angles of Proposition 1 and the preceding paragraphs. Left image: A rotation by an angle $\bar{{\alpha}}$ in the $X-Y$ plane with the $Z-$axis fixed. Centre image: A rotation by an angle $\gamma$ in the $X'-Z'$ plane with the $Y'-$axis fixed. Right image: A rotation by an angle $\bar{\beta}$ in the $X''-Y''$ plane with the $Z''-$axis fixed.}
\label{fig:Euler-angles}
\end{figure}

Three successive rigid rotations may be applied that bring $\vec{p}$ to $\vec{p}\,'$. These are not uniquely defined, but applying the following conventions yields uniqueness: (i) the first rotation is about the $Z$ axis, rotating by an angle $\bar{\alpha}\in(-\pi,\pi]$; (ii) the second is  about the $Y'$ axis (i.e.\ the new $Y$ axis after rotation (i)) and has magnitude $\gamma\in [0,\pi)]$  (the geodesic distance/arc length from $P$ to $P'$); and (iii) the third is about the $Z''$ axis (i.e.\ the new $Z$ axis after rotations (i) and (ii)), rotating by an angle $\bar{\beta}\in(-\pi,\pi]$. This canonical choice yields uniquely defined Euler angles $(\bar{\alpha},\gamma,\bar{\beta})$. See Fig.~\ref{fig:Euler-angles}. This is a Type $A$ rotation in the language of \cite{varshalovich1988quantum}, often referred to as a $Z-Y-Z$ rotation (\cite{varshalovich1988quantum}, p.\ 22). 
Then (see \cite{varshalovich1988quantum,monteverdi2024some,michel2020mathematical})
\begin{eqnarray}
    \cot\bar{\alpha} &=& \cos\theta\cot(\phi-\phi')-\cot\theta'\sin\theta\csc(\phi-\phi'),\label{albar-angles} \\
    \cot\bar{\beta} &=& \cos\theta'\cot(\phi-\phi')-\cot\theta\sin\theta'\csc(\phi-\phi'). \label{bebar-angles}
\end{eqnarray}
These formulae are not valid when $\phi-\phi'$ is an integer multiple of $\pi$. However, this will not affect the integrals over $\mathbb{S}^2$ where $\Delta\phi=\phi-\phi'=\pi$ only at endpoints of these integrals - and where any singularity is of integrable form.

We now return to (\ref{uring-transport}). In  the spin-0 case, 
\be \mathring{U}=\left.{}_s\mathring{U}\right|_{s=0}= \left(\frac{\gamma}{\sin\gamma}\right)^{1/2}, \label{vv-s2}
 \ee
 so that ${{\mathring{U}}}$ is the square root of the van Vleck determinant on $\mathbb{S}^2$ \cite{allen1986vector}.
 This satisfies (\ref{uring-transport}) with $s=0$. Then we consider the ansatz (see Eq.~(89) in~\cite{2023PhRvD.108l5017I} for a similar approach in the 4D case) 
 \be {_s{\mathring{U}}}(x^A,x^{A'})=e^{\zeta_s(x^A,x^{A'})} \mathring{U}(x^A,x^{A'}),\label{zeta-def} \ee
 for some function $\zeta_s(x^A,x^{A'})$.
 Note that since $[{_s{\mathring{U}}}]=[\mathring{U}]=1$, we must have 
 \be [\zeta_s]=0.\label{zeta-initial} \ee
 Inserting this into (\ref{uring-transport}) with $s=0$ yields 
 \be 2\mathring{\sigma}^A\partial_A\zeta_s+a^A\partial_A\mathring{\sigma} = 0. \label{zeta-transport1} \ee
 Using 
 \be \mathring{\sigma}=\frac12\gamma^2, \label{sigma-gamma} \ee
 and (from (\ref{A-vec}))  \be a^A\partial_A = 2is\frac{\cos\theta}{\sin^2\theta}\partial_\phi, \label{avec-s2} \ee
 we can write (\ref{zeta-transport1}) as 
 \be
     (-\sin\theta\cos\theta'+\cos\theta\sin\theta'\cos(\phi-\phi'))\partial_\theta\zeta_s -\csc\theta\sin\theta'\sin(\phi-\phi')\partial_\phi\zeta_s 
     -is\cos\theta\csc\theta\sin\theta'\sin(\phi-\phi') =0. \label{zeta-transport2} 
 \ee
 This is a transport equation along geodesics of $\mathbb{S}^2$, and to solve it we use a coordinate system for which the initial point of the geodesic lies on the equator. So, without loss of generality, we can take $(\theta'=\pi/2,\phi'=0)$. In this case the geodesic distance is given by 
 \be \cos\gamma = \sin\theta\cos\phi,\label{gamma-special} \ee
 and the transport equation (\ref{zeta-transport2}) reads
 \be \cos\theta\cos\phi\,\partial_\theta\zeta_s-\csc\theta\sin\phi\,\partial_\phi\zeta_s=is\cot\theta\sin\phi. \label{zeta-transport3} \ee
 The characteristics of this equation (that is, the geodesics) satisfy 
 \be \frac{d\phi}{d\theta} = -\frac{\tan\phi}{\cos\theta\sin\theta}. \label{s2-chars} \ee
 Integrating yields 
 \be \sin\phi = \kappa\cot\theta \label{kap-def} \ee
 for a quantity $\kappa$ that is constant along each individual characteristic $C$.
 Then (\ref{zeta-transport3}) can be written as an ODE along characteristics: 
 \be \frac{d}{d\theta}\left(\left.\zeta_s\right|_{C}\right) = \left.is\csc\theta\tan\phi\right|_{C}\label{zeta-transport4} \ee
 Using (\ref{kap-def}) to substitute for $\phi$, we can integrate this equation. We assume for the moment that $\phi\in (0,\pi/2)$. From (\ref{gamma-special}), this gives $\cos\gamma>0$, and the relevant sign choices in the calculations below are consistent. We apply the initial condition $\zeta_s=0$ when $\theta=\pi/2$ (which corresponds to (\ref{zeta-initial}) and also yields the initial condition $\phi=0$) to obtain 
 \be \left.\zeta_s\right|_C = is \left(\arctan\left(\frac{1}{\kappa}((1+\kappa^2)\sin^2\theta-\kappa^2)^{1/2}\right) -\arctan\left(\frac{1}{\kappa}\right)\right).\label{zeta-sol1} \ee
To solve (\ref{zeta-transport2}), we now step off the characteristics by substituting for $\kappa$ from (\ref{kap-def}). We find
\be ((1+\kappa^2)\sin^2\theta-\kappa^2)^{1/2}=\sin\theta\cos\phi = \cos\gamma,\label{key-term-gamma} \ee
and so we can write 

\be \zeta_s = {is}\left( \arctan\left(\frac{\cos\gamma}{\kappa}\right) -\arctan\left(\frac{1}{\kappa}\right) \right).\label{zeta-sol2} \ee

Recall that we are on a geodesic with initial point $(\theta'=\pi/2,\phi'=0)$. In this case, the Euler angles $\bar{\alpha}$ and $\bar{\beta}$ of (\ref{albar-angles}) and (\ref{bebar-angles}) satisfy 
\begin{eqnarray} 
\cot\bar{\alpha} &=& \cos\theta\cot\phi, \label{albar-special}\\
\cot\bar{\beta} &=& -\cot\theta\csc\phi. \label{bebar-special} 
\end{eqnarray} 
Then we see, using Eqs. (\ref{gamma-special}) and (\ref{kap-def}),  that the arguments of the $\arctan$'s in (\ref{zeta-sol2}) are essentially Euler angles: 
\be \frac{\cos\gamma}{\kappa}=\cot\bar{\alpha},\quad \frac{1}{\kappa} = -\cot\bar{\beta}.\label{Euler-subs} \ee
Depending on the sign of $\kappa$  (which is non-zero  and finite), one of $\cot\bar{\alpha},\cot\bar{\beta}$ is positive, and the other negative (recall that $\cos\gamma>0$ in this case). Using the identity $\cot(C+\pi)=\cot(C)$, we can assume without loss of generality that the angle $A$ (one of $\bar{\alpha}, \bar{\beta}$) with $\cot(A)>0$ lies in $(0,\pi/2)$, and that the angle $B$ (the other of $\bar{\alpha}, \bar{\beta}$) with $\cot(B)<0$ lies in $(-\pi/2,0)$.  Then we have 
\be \arctan(\cot(A))=\frac{\pi}{2}-A,\quad \arctan(\cot(B)) = -\frac{\pi}{2}-B. \label{arccot-rules} \ee and so in both cases, (\ref{zeta-sol2}) yields 
\be \zeta_s = -is(\bar{\alpha}+\bar{\beta}). \label{zeta-final} \ee

We now drop the restrictions of on $\phi,\gamma,\bar{\alpha}$ and $\bar{\beta}$ and use a direct calculation to show that this formula holds in all cases. This yields an expression for $\zeta_s$ that is purely geometric in the sense that it involves only the Euler angles $\bar{\alpha}$ and $\bar{\beta}$ that are invariantly defined as above. Thus the choice of coordinate system that places the initial point at $(\theta',\phi')=(\pi/2,0)$ is not relevant, and (\ref{zeta-final}) holds generally.

Thus we have the following formula for ${_s{\mathring{U}}}$, the $\mathbb{S}^2$-factor of the direct part ${_s{\hat{U}}}$ of the Hadamard form of the retarded Green function in $\mathcal{M}_2\times\mathbb{S}^2$: 

\be {_s{\mathring{U}}} = e^{-is(\bar{\alpha}+\bar{\beta})}\left(\frac{\gamma}{\sin\gamma}\right)^{1/2}, \label{uring-sol} \ee
where the Euler angles $\bar{\alpha}$ and $\bar{\beta}$ are given via (\ref{albar-angles}) and (\ref{bebar-angles}) respectively. 

\subsubsection{An aside on Euler angles.}
There is an interesting corollary of the calculation above which we state as the following proposition. This relates geodesics and Euler angles on $\mathbb{S}^2$. The result follows immediately from the fact that $\kappa$ is constant along geodesics, and from the formula (\ref{Euler-subs}). 

\begin{proposition}
    Let $C$ be a great circle of $\mathbb{S}^2$ and fix $P'\in C$. Let $P$ be any other point of $C$, and let $\bar{\alpha}, \gamma$ and $\bar{\beta}$ be the Euler angles of the $Z-Y-Z$ (Type A) rotation that takes the position vector $\vec{OP}'$ to $\vec{OP}$, where $O$ is the origin of $\real^3$ and $\gamma$ is the arc length on $\mathbb{S}^2$ between $\vec{OP}'$ on $\vec{OP}$. Then $\bar{\beta}$ is independent of the choice of $P$. That is, the third Euler angle is the same for all points on a given geodesic through $P'$. 
\end{proposition}

\subsection{The direct part on $\mathcal{M}_2$.}\label{sec:direct-M2}

We now turn to the direct part ${_s{\bar{U}}}$ on $\mathcal{M}_2$. This satisfies the transport equation (\ref{ubar-transport}) with the coincidence limit $[{_s{\bar{U}}}]=1$. We adopt the same approach as we did for ${_s{\mathring{U}}}$. That is, we use the ansatz
\be 
{_s{\bar{U}}}{(x^i,x^{i'})} = e^{\lambda_s{(x^i,x^{i'})}} {\bar{U}}{(x^i,x^{i'})},\label{lambda-def} \ee
for some function $\lambda_s(x^i,x^{i'})$ and where 
\be {\bar{U}} = \left.{_s{\bar{U}}}\right|_{s=0}.\label{ubar-0} \ee

This is the $\mathcal{M}_2$ factor of ${_s{\hat{U}}}$ for the spin-$0$ case. Unlike the case for the corresponding $\mathbb{S}^2$ factor, we do not have an explicit formula for ${\bar{U}}$. However, it satisfies the transport equation (\ref{ubar-transport}) with $s=0$ and the coincidence limit $[{\bar{U}}]=1$. Then $\lambda_s$ must satisfy the initial value problem 
\be \bar{\sigma}^i\partial_i\lambda_s + sa^i\partial_i\bar{\sigma}=0,\quad [\lambda_s]=0. \label{lambda-transport} \ee

We can solve this transport equation (and the transport equation for ${\bar{U}}$) as follows. Note that the transport is along a geodesic of $(\mathcal{M}_2,\bar{g})$. See \cite{nolan2026spray} for details. We fix a point $p_0:(t,r)=(t_0,r_0)$ of $\mathcal{M}_2$ and introduce coordinates $(\tau,\xi)$  on the chronological future $\cd=I^+(p_0)$ of $p_0$ as follows\footnote{$p_0$ corresponds to the base point with local coordinates $x^{i'}$, and $p$ to the field point with local coordinates $x^i$. With an abuse of notation, we will use $Q(x^i,x^{i'})$, $Q(p_0,p)$ and $Q(\tau,\xi)$ interchangeably for 2-point functions $Q$ on $\mathcal{M}_2$. The coordinates $(\tau,\xi)$ are defined in (\ref{tau-def}) and (\ref{xi-def}) respectively.}. Recall that $(\mathcal{M}_2,\bar{g})$ is a causal domain, and so all pairs of points $(p,q)\in \mathcal{M}_2{\times \mathcal{M}_2}$ are connected by a unique geodesic. For $p\in\cd$, we define
\be \left.\tau\right|_p =  \hbox{the proper time at $p$ along the unique timelike geodesic from $p_0$ to $p$} \label{tau-def} \ee
and 
\be \left.\xi\right|_p = \hbox{$\dot{r}(0)$ on this geodesic.} \label{xi-def} \ee
We use an overdot for the derivative with respect to proper time $\tau$ along the geodesic. Note that $\xi$ is constant along each individual timelike geodesic from $p_0$. Then, as shown in \cite{nolan2026spray}, we can write the line element of $\mathcal{M}_2$ as 
\be ds^2 = -d\tau^2 + h^2(\tau,\xi)d\xi^2,\label{spray-metric} \ee
where $h$ is given explicitly in terms of (elliptic) integrals along geodesics. In these coordinates, we have 
\be \bar{\sigma} = -\frac12\tau^2, \label{sigma-tau} \ee
and the solution of the transport equation (\ref{lambda-transport}) can be written as 
\be \lambda_s(\tau,\xi) = s\int_0^\tau \left((\partial_t\tau')a^t+(\partial_r\tau')a^r \right) d\tau'. \label{lambda-transport-sol} \ee
The integrand here contains the explicitly known components of $a^i\partial_i$, and components of the inverse Jacobian matrix of the coordinate transformation $(t,r)\mapsto (\tau,\xi)$ \cite{nolan2026spray}. We outline briefly how this integral can be calculated (along with the corresponding integral for $\bar{U}$) in the case of a worldline of Schwarzschild along a constant radius $r_0>3M$. Note that this includes, in particular, the relevant case of a circular timelike geodesic of Schwarzschild spacetime at the radius $r_0$. 

In this case, the 4-dimensional worldline projects onto a curve of $\mathcal{M}_2$ on which $r$ is constant, so with $p_0:(t,r)=(t_0,r_0)$, our aim is to calculate ${_s{\bar{U}}}(p_0,p)$ with $p:(t,r)=(t_0+\Delta t, r_0)$. It can be shown (see~\cite{casals2015geometric}) that the relevant timelike geodesic of $(\mathcal{M}_2,\bar{g})$ from $p_0$ to $p$ is initially ingoing, reaches a minimum of $r$ at $r_{\rm{m}}\in(3M,r_0)$, and is then outgoing, returning to $r=r_0$. This corresponds to values of $\xi$ with $\xi\in(-\xi_c,0)$ where $\xi_c>0$ is referred to as the critical value: the geodesic from $p_0$ with $\dot{r}(0)=-\xi_c$ is future complete, and satisfies $\lim_{\tau\to+\infty}r(\tau)=3M$ \cite{casals2015geometric}. The metric function $h(\tau,\xi)$ takes a different form along the ingoing and outgoing branches of the geodesic, and so (\ref{lambda-transport-sol}) must be written as 
\be \lambda_s(\tau,\xi) = s\left(\int_0^{\tau_1(\xi)} \cdots d\tau' + \int_{\tau_1(\xi)}^\tau \cdots d\tau'\right) \label{lambda-int-split} \ee
where $\tau_1(\xi)$ is the value of proper time at the minimum $r_{\rm{m}}=r_{\rm{m}}(\xi)$ of the geodesic labelled by $\xi$. 
The timelike geodesics on $\mathcal{M}_2$ satisfy~\cite{casals2015geometric}
\be \dot{t} = E\frac{r^2}{f},\quad \dot{r}^2 = R(E,r), \label{M2-geos} \ee
where $E$ is the conserved energy along the geodesic and 
\be R = E^2r^4-r^2f. \label{R-def} \ee
Since $r_0$ is the initial value of $r$ on the geodesic, we have $E=E(\xi)$ with (\textit{cf.} (\ref{xi-def}))
\be \xi^2 = E^2r_0^4-r_0^2f(r_0). \label{xi-ep} \ee
$\xi\in(-\xi_c,0)$ is the negative root of this equation, and the corresponding range for $\ep$ is $\ep\in(0,\ep_c)$ where $\ep_c =1/(3\sqrt{3}M)$. The minimum radius on the geodesic is $r_{\rm{m}}=r_{\rm{m}}(\xi)\in(3M,r_0)$ where $r_{\rm{m}}$ is the larger of the two positive roots of $R(\ep,r)=0$. Due to the static nature of the spacetime, the proper time $\tau_1$ and coordinate time $t_1$ from $r_0$ to $r_{\rm{m}}$ (ingoing) {are} the same as, respectively, the proper and coordinate times from $r_{\rm{m}}$ to $r_0$ (outgoing) and we can calculate 
\begin{eqnarray} \tau(p_0,p) &=&  2\,\tau_1(\xi) = 2\int_{r_{\rm{m}}(\xi)}^{r_0} \frac{dr}{R^{1/2}(\ep(\xi),r)}, \label{tau-in-out} \\
\Delta t(p_0,p) &=& 2\,t_1(\xi) = 2\ep \int_{r_{\rm{m}}(\xi)}^{r_0} \frac{r^2}{fR^{1/2}(\ep(\xi),r)}dr. \label{t-in-out} 
\end{eqnarray}
Then we can show that 
the van Vleck determinant in $\mathcal{M}_2$ at $p$ has the value (\cite{nolan2026spray})
\be \bar{\Delta}(p_0,p) = \frac{\tau}{2\xi\tau_1'(\xi)}=\frac{\tau_1(\xi)}{\xi\tau_1'(\xi)}. \label{vv-sol} \ee
Comparing (\ref{ubar-transport}) with $s=0$ with the transport equation satisfied by the van Vleck determinant $\bar{\Delta}$ (see e.g.\ \cite{poisson2011motion}) shows that, in general (not just for constant radius $r$),
\eqn{\bar{U}{(x^i,x^{i'})}=\bar{\Delta}^{1/2}{(x^i,x^{i'})}.\label{eq:bar-U}} 
We can calculate the relevant inverse Jacobi matrix elements \cite{nolan2026spray}, and we find 
\be \lambda_s(p_0,p) = 2s\lambda(p_0,p),\quad \lambda(p_0,p)= -\int_{r_{\rm{m}}}^{r_0} \frac{\ep\, r}{fR^{1/2}}\left(1-\frac{3M}{r}\right) dr.
\label{lambda-final} \ee
Monotonicity of $r$ on the separate branches of (\ref{lambda-int-split}) allows us to take $r$ as the integration variable. A useful simplification takes place  - the terms with $a^t$ are equal on each branch and the terms with $a^r$ are equal in magnitude and opposite in sign - leaving (\ref{lambda-final}).

We plot {the} results \eqref{tau-in-out}, \eqref{vv-sol} and \eqref{lambda-final} for, respectively, $\tau$, $\bar\Delta$ and $\lambda$ by treating the relevant integrals as parametric functions of $r_{\rm{m}}\in(3M,r_0]$. Smaller values of $r_{\rm{m}}$ correspond to later times: the relevant geodesics penetrate deeper into the potential well, and return to $r_0$ at later times, and $r_{\rm{m}}\to 3M$ corresponds to $\Delta t\to+\infty$. $r_{\rm{m}}=r_0$ corresponds to coincidence. $\ep$ (and hence $\xi$) is determined by noting that $\ep^2r_{\rm{m}}^2-f(r_{\rm{m}})=0$. Some manipulations are required to evaluate the derivative $\tau_1'(\xi)$  in \eqref{vv-sol}: the fundamental theorem of calculus cannot be applied directly as the integrand is singular at $r=r_{\rm{m}}$. Integration by parts allows us to deal with this singular behaviour. Likewise, the integrals of (\ref{tau-in-out}), (\ref{t-in-out}) and (\ref{lambda-final}) are numerically unstable  in the coincidence limit $r_{\rm{m}}\to r_0$. We can account for this by making the following change of variables (the integrals mentioned have the form below for functions $H$ that are analytic in $r_{\rm{m}}$ at $r_{\rm{m}}=r_0$): 
\be \int_{r_{\rm{m}}}^{r_0} \frac{dr}{(r-r_{\rm{m}})^{1/2}H(r,r_{\rm{m}})} = \int_0^{(r_0-r_{\rm{m}})^{1/2}} \frac{2}{\tilde{H}(u,r_{\rm{m}})}du \label{integrate-in-u} \ee
where 
\be u=(r-r_{\rm{m}})^{1/2},\quad \tilde{H}(u,r_{\rm{m}})=H(r_{\rm{m}}+u^2,r_{\rm{m}}).\label{u-def} \ee
See App.~\ref{sec:AppA} and \cite{nolan2026spray} for details on the mollification of the integrals  carried out as in Eq.~\eqref{integrate-in-u}. 

See App.~\ref{sec:AppB} for expressions of $\tau_1$, $\Delta t$, $\lambda$ and $\bar\Delta$ in terms of elliptic integrals, obtained by re-writing the integral representations above for these quantities.

\begin{figure}[h!]
\begin{center}
\includegraphics[width=8cm]{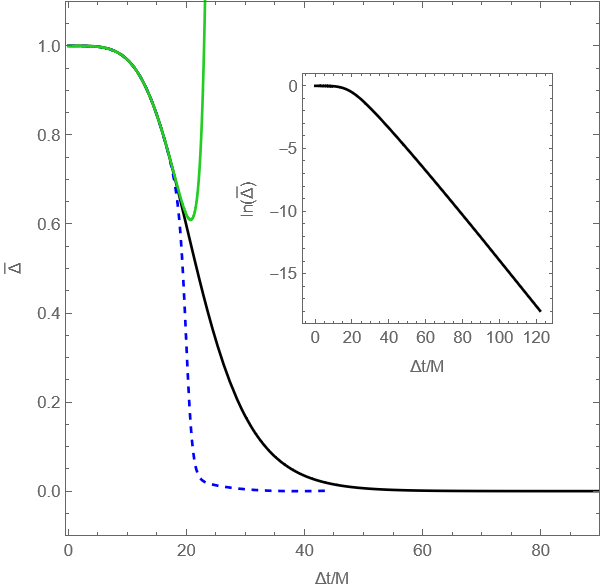}
  \end{center}
\caption{Plot of the van Vleck determinant $\bar\Delta$ of $\mathcal{M}_2$ as a function of coordinate time $\Delta t$ along a constant radius trajectory of Schwarzschild spacetime at radius $r=6M$. The plot shows the results of three different calculations. Solid green: small coordinate expansion to $\mathcal(\Delta x)^{20}$. Dashed blue: near-coincidence expansion to $\mathcal(r_{\rm{m}}-r_0)^{12}$. See Appendix \ref{sec:SmallCoordExpansionForBarU} for details of these expansions, and their comparison. Black: direct evaluation of the elliptic integrals of Appendix \ref{sec:AppB}. The small coordinate expansion plots $\bar\Delta$ directly as a function of $\Delta t$: the other plots are parametric plots of $\bar\Delta(r_{\rm{m}})$ against $\Delta t(r_{\rm{m}})$. The inset shows the plot of the logarithm of $\bar\Delta$ for the elliptic integral method, and indicates exponential decay of $\bar\Delta$ for large $\Delta t$. The results are consistent with Fig. 6(b) of \cite{casals2015geometric}. The relevant calculations and plots were carried out using Mathematica.}
\label{fig:vv6M}
\end{figure}

As indicated, we have different ways to evaluate (and plot) the expressions obtained for $\tau, \bar{\Delta}, \lambda$ and ${}_s\bar{U}$. We have the mollified integrals of Appendix A and the elliptic integrals of Appendix B. The plots are generated using Mathematica, and we find that plotting the mollified integrals of Appendix \ref{sec:AppA} yields results indistinguishable from the plots of the elliptic integrals of Appendix \ref{sec:AppB}. For this reason, we omit the former (but as we will see, the calculations of Appendix \ref{sec:AppA} are used for another purpose). We can also calculate these quantities in an entirely different way by determining small coordinate distance expansions. This was done for $\bar\sigma$ and $\bar{\Delta}$ in~\cite{CNOW2019} (see also \cite{aruquipa2026greenfunctionsreggewheelerteukolsky}). In  Appendix \ref{sec:SmallCoordExpansionForBarU} we reproduce these results for $\bar\sigma$ and we  extend them to ${}_s\bar{U}$. We can connect the methods by expanding the integrals of Appendix A in powers of $\Delta t$.  This allows us to validate our results by, on the one hand, comparing numerical plots, and on the other by evaluating expansion coefficients in two independent ways. The comparison of expansion coefficients is carried out in Appendix \ref{sec:SmallCoordExpansionForBarU}. The numerical comparison is summarised in Figure \ref{fig:vv6M}, where we plot the van Vleck determinant $\bar{\Delta}$ of $(\mathcal{M}_2,\bar{g})$ along a timelike worldline  at fixed radius using (in  all) four different approaches. As we see, we have excellent agreement over a time interval up to $\Delta t\sim 20M$. We shall see later that this actually covers almost the entirety of the relevant time range. The exact calculation yielding elliptic integral representations covers the entire range.

\begin{figure}
    \includegraphics[width=8cm]{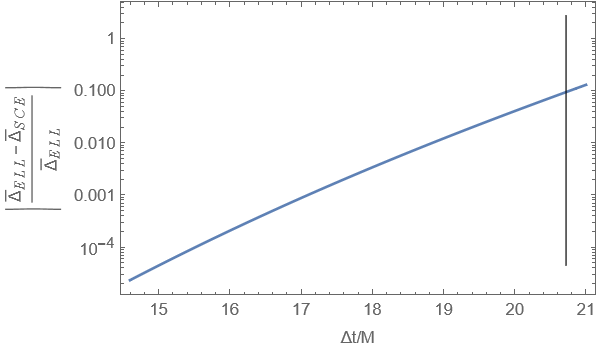}
    \caption{Log-plot of the relative error in the setting of Fig.~\ref{fig:vv6M} in the values of $\bar{\Delta}_{\textrm{SCE}}$ - the van Vleck determinant calculated using the small coordinate expansion of Appendix \ref{sec:SmallCoordExpansionForBarU} - compared to the exact value $\bar{\Delta}_{\textrm{ELL}}$ calculated using the elliptic integrals of Appendix \ref{sec:AppB}. The plot also shows the key value  $\Delta t\simeq 20.72 M$ (see the discussion of Section \ref{sec:ell-modes}). The relative error is below 10\% up until this time.}
    \label{fig:error}
\end{figure}

\begin{figure}[h!]
\begin{center}
\includegraphics[width=6cm]{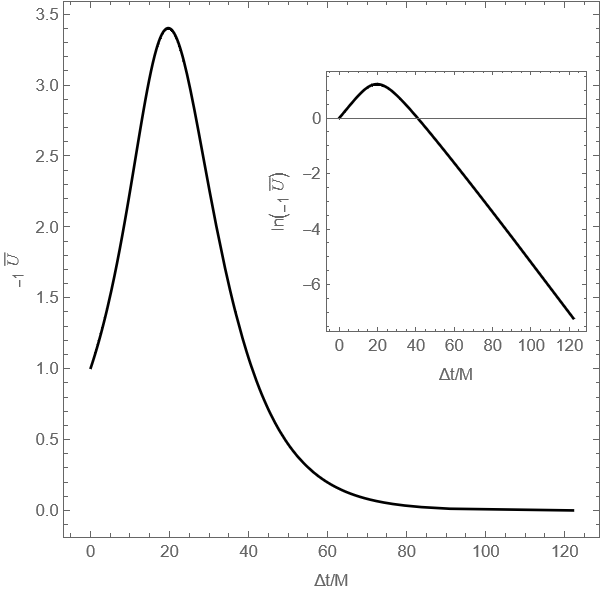}
\includegraphics[width=6cm]{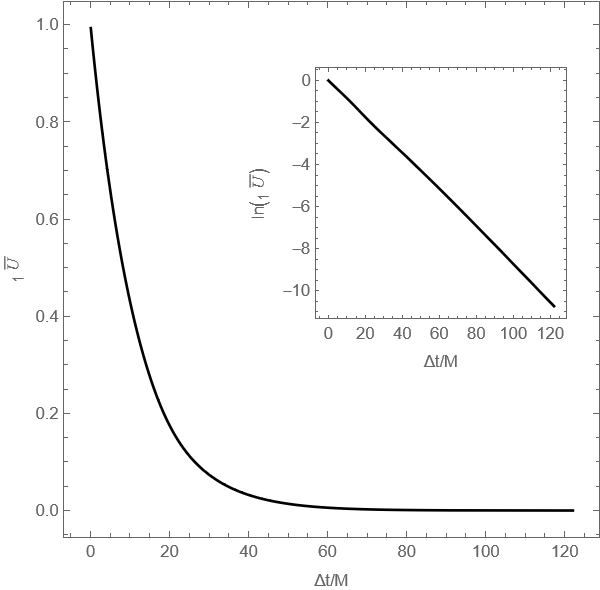}
\includegraphics[width=6cm]{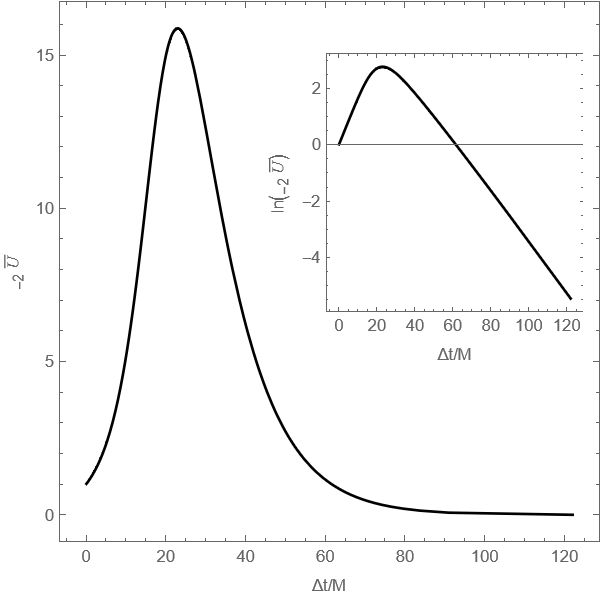}
\includegraphics[width=6cm]{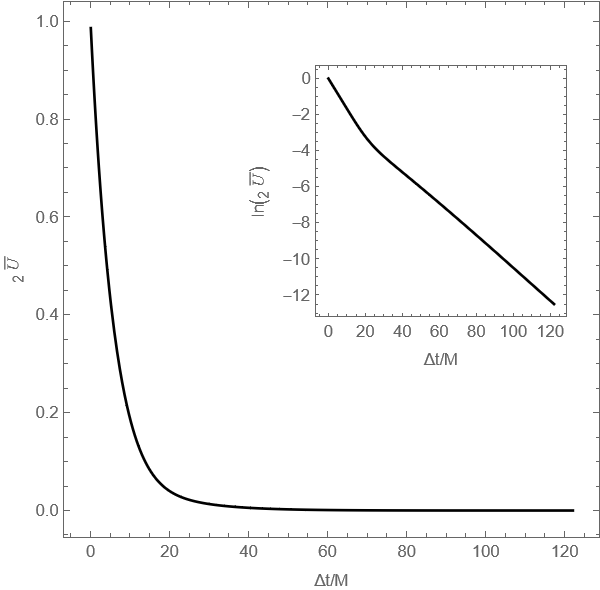}
\end{center}
\caption{The  coefficient ${}_s\bar{U}$ in the direct part of the BPT GF for $s=\pm1$ (top) and $s=\pm2$ (bottom) along a 
trajectory of Schwarzschild spacetime with constant radius $r = 6M$. The figures show parametric plots of ${}_s\bar{U}(r_{\rm{m}})$ against $\Delta t(r_{\rm{m}})$ for $r_{\rm{m}}\in(3M,r_0)$ with $r_0=6M$.  In each case, the insets show exponential decay of ${}_s\bar{U}$ as $\Delta t$ increases.}
\label{fig:sU}
\end{figure}

\begin{figure}[h!]
\begin{center}
\includegraphics[width=6cm]{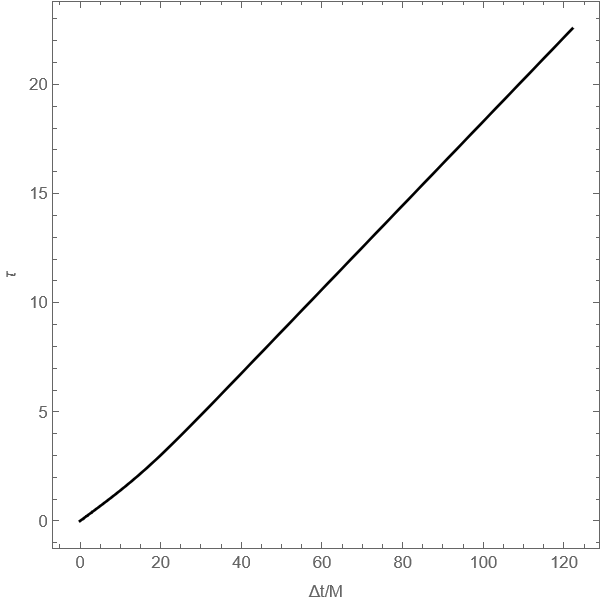}
\includegraphics[width=6cm]{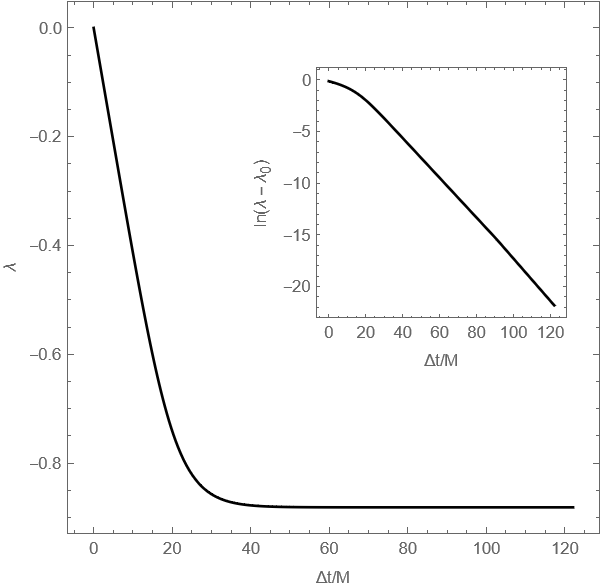}
\end{center}
\caption{Left image: Plot of the proper time $\tau$ along the constant radius trajectory at $r=6M$.   Right image: Plot of the BPT parameter $\lambda$ and (inset) $\ln(\lambda-\lambda_0)$, indicating the exponential decay as $\Delta t\to+\infty$ of $\lambda$ to $\lambda_0\simeq -0.881374$. In both cases, the plots are parametric plots with parameter $r_{\rm{m}}$ of the elliptic integrals of Appendix~\ref{sec:AppB}.}
\label{fig:tau-lam}
\end{figure}

The small coordinate  and the near-coincidence expansions agree analytically (as noted in in Appendix \ref{sec:SmallCoordExpansionForBarU}), and so the difference in their curves in Figure \ref{fig:vv6M} is purely due to plotting them to different orders (not because they are different expansion methods).
Figure \ref{fig:vv6M} indicates that the small coordinate expansion and near-coincidence expansions are valid (as quantified by the relative error plotted in Fig.~\ref{fig:error}) only up to $\Delta t\sim 20M$. As noted, the expressions yielding the modified integral and elliptic integral plots are exact, and the resulting plots are in full agreement: only the latter are shown. These observations in hand, we provide further plots only for the elliptic integrals. Figure \ref{fig:sU} shows plots of the coefficient ${}_s\bar{U}$ of the direct Hadamard term in the BPT GF for $s=\pm1$ and $s=\pm2$ along the worldline of constant radius $r=6M$. In each case, the coefficients ultimately undergo exponential decay. For negative $s$, there is an initial period of growth, not present in the spin-0 (or $s>0$) case. Figure \ref{fig:tau-lam} shows plots of the proper time $\tau$ and the Teukolsky parameter $\lambda$ along the same worldline.
We checked that the plot of $\tau$ in the left of  Figure \ref{fig:tau-lam} is consistent  with the plot of~\footnote{In Figure 6a of~\cite{casals2015geometric}, two of the authors said that $\bar\sigma$ (there denoted by ``$\sigma_2$") was being plotted but that was a typographical error: it was instead $\bar\sigma/4$ which was being plotted.} $\bar\sigma/4=-\tau^2/8$ in Figure 6a of~\cite{casals2015geometric}.
On the other hand, the plot of $\lambda$ in the right of  Figure \ref{fig:tau-lam} is new.

\section{Non-direct part of the BPT GF}\label{sec:non-direct}

The main purpose of this section is to calculate the non-direct part of the BPT GF, which is defined by subtracting the  direct Hadamard part $\Gd{s}(x,x'):={}_sU(x,x')\delta_+(\sigma)$ in \eqref{eqn:hadamardTGret} from the full BPT GF:
\begin{equation}\label{eq:Gnd Had}
\Gnd{s}(x,x') := \left\{\begin{array}{l l}
\Gret{s}(x,x')-\Gd{s}(x,x'), & x'\in \mathcal{N}_\text{max}(x);  \\
\displaystyle
\Gret{s}(x,x'),&x'\notin \mathcal{N}_\text{max}(x),
\end{array}
\right.
\end{equation}
where $\mathcal{N}_\text{max}(x)$refers to the {\it maximal} normal neighbourhood of $x$.
Now, using the Hadamard form \eqref{eqn:hadamardTGret} of the GF, we can write 
\be 
\Gret{s}(x,x')-\Gd{s}(x,x')=\Gret{s}(x,x')-{}_sU(x,x')\delta(\sigma)\theta(\dt)=
{}_sV(x,x')\theta(-\sigma)\theta(\dt), 
\quad x'\in \mathcal{N}(x).
\ee

In order to carry out the subtraction of the direct part as per \eqref{eq:Gnd Had}, we shall decompose both the GF and its direct part 
 into  spin-weighted spherical harmonics (SWSH) $\sYlm(\theta,\phi)$~\cite{goldberg1967spin}, which satisfy 
 \be {}_s\mathcal{L}\left(\sYlm(\theta,\phi)\right) = -(\ell-s)(\ell+s+1)\sYlm(\theta,\phi),\qquad \ell \geq|s|,\quad -\ell\leq m \leq \ell. \label{sYlm-eigenfunctions} \ee
 The operator ${}_s\mathcal{L}$ is given in \eqref{conf-split-s2}. For the BPT GF, this decomposition reads
\be
    \s G_\textrm{ret}=4\pi \lP\Delta_S(r')\rP^s\sum\limits_{\ell=|s|}^{\infty}\sum\limits_{m=-\ell}^{\ell}\sGTl{s}\lP r,r',\Dt\rP\sYlm(\theta,\phi)\sYlm^*(\theta',\phi'),
\label{sGret-decomp} \ee
where $\sGTl{s}$ are the $\ell$-modes of the BPT GF  
and $\Delta_S(r'):=r'(r'-2M)$.
 We choose the SWSH with a normalization so that their orthonormality relation is
    \be\label{eq:norm-SWSH}
\int_{\mathbb{S}^2}
\sYlm(\theta,\phi)\,\s Y^*_{\ell',m'}(\theta,\phi)\df\Omega=\delta_{\ell\ell'}\delta_{mm'}
\ee
and their completeness relation is
\eqn{\label{eq:Compl.Rln}
\sum_{\ell=|s|}^{\infty}\sum_{m=-\ell}^{\ell}\sYlm(\theta,\phi)\,\s Y^*_{\ell m}(\theta',\phi')=\delta(\cos\theta-\cos\theta')\delta(\phi-\phi').
}
Using the addition  law~\cite{michel2020mathematical}
\begin{equation} \label{eq:add SWSH}
\sum\limits_{m=-\ell}^{\ell}\sYlm(\theta,\phi)\sYlm^*(\theta',\phi')=
(-1)^{s}\sqrt{\frac{2\ell+1}{4\pi}}e^{-i s\bar\alpha}\s Y_{\ell,-s}\lP\gamma,\bar\beta\rP,
\end{equation}
we can re-write (\ref{sGret-decomp}) in terms of one single ($\ell$-)sum:
\eqn{
\s G_\textrm{ret}(x,x')=
\sqrt{4\pi} e^{-i s(\bar\alpha+\pi)}\lP\Delta_S(r')\rP^s  \sum\limits_{\ell=|s|}^{\infty}\sqrt{2\ell+1}\,\sGTl{s}\lP r,r',\dt\rP\,\s Y_{\ell,-s}\lP\gamma,\bar\beta\rP.\label{eq:mode-dec-GT}}
 We note that $\bar{\alpha}$ and $\bar{\beta}$ are the Euler angles of (\ref{albar-angles}) and (\ref{bebar-angles}) respectively.
 As is well-known (see, e.g.,~\cite{casals2023global,aruquipa2026greenfunctionsreggewheelerteukolsky} in the cases of the scalar equation and of the related Regge-Wheeler equation), the singular term of $\s G_\textrm{ret}$
arises from Eq.~\eqref{eq:mode-dec-GT} as a divergence in
the $\ell$-sum when considered as a classical function (the sum converges to a distribution, i.e.\ it converges weakly).

We can carry out a similar $\ell$-mode decomposition of the direct part. We define the mode functions $ \sGTld{s}\lP r,r',\dt\rP$ as the coefficients in the following expansion in terms of SWSHs:
\begin{align}
\Gd{s}(x,x')&=
\frac{1}{r\cdot r'}
    \sum\limits_{\ell=|s|}^{\infty}
    \sum\limits_{m=-\ell}^{\ell}
    \sGTld{s}\lP r,r',\dt\rP\,
    \sYlm(\theta,\phi)\sYlm^*(\theta',\phi')
   \nonumber \\ &
=
\frac{(-e^{-i\bar\alpha})^{s}}{\sqrt{4\pi}\,r\cdot r'}   \sum\limits_{\ell=|s|}^{\infty}\sqrt{2\ell+1}\,\sGTld{s}\lP r,r',\dt\rP\,\s Y_{\ell,-s}\lP\gamma,\bar\beta\rP,
\label{eq:mode-dec-GTd}
    \end{align}
 where in the last equality we have again used Eq.~\eqref{eq:add SWSH} and we refer to $\sGTld{s}$ as the $\ell$-modes of the direct part of the GF.

Using the definition \eqref{eq:Gnd Had} of the  non-direct part $\Gnd{s}$ of the GF, together with the $\ell$-mode decompositions \eqref{eq:mode-dec-GT} and \eqref{eq:mode-dec-GTd} of, respectively the GF and its direct part, we readily obtain the following $\ell$-mode decomposition of the non-direct part:
\begin{eqnarray}
\Gnd{s}&=&4\pi\lP\Delta_S(r')\rP^{s}\sum\limits_{\ell=|s|}^{\infty}\lP\sGTl{s}-\sGTldt{s}\rP\sum\limits_{m=-\ell}^{\ell}\sYlm(\theta,\phi)\sYlm^*(\theta',\phi')
  \nonumber  \\
   &=& \sqrt{4\pi}\lP-e^{-i \bar\alpha}\Delta_S(r')\rP^s  \sum\limits_{\ell=|s|}^{\infty}\sqrt{2\ell+1}\lP\sGTl{s}-\sGTldt{s}\rP\,\s Y_{\ell,-s}\lP\gamma,\bar\beta\rP \label{eq:decomp-nondirect}
\end{eqnarray}
where\footnote{We note that, for $s=0$, our definition of $\sGTld{s}$ is equal to $G_{\ell}^{\rm d}$ in Eq.~(19) in  \cite{CNOW2019} multiplied by $4\pi$.}
\eqn{
    \sGTldt{s}:=\frac{\left(\Delta_S(r')\right)^{-s}}{4\pi r r'}\sGTld{s}.
}

In the next subsection we shall calculate the $\ell$-modes $\sGTld{s}$ of the direct part, while in~\cite{aruquipa2026greenfunctionsreggewheelerteukolsky} two of the authors presented ways for calculating the $\ell$-modes $\sGTl{s}$ of the full GF.
In the following subsection, we shall use the calculation of $\sGTld{s}$ and  $\sGTl{s}$ in order to calculate the non-direct part of the GF via  Eq.~\eqref{eq:decomp-nondirect}.

\subsection{$\ell$-modes of the direct part of the BPT GF}\label{sec:ell-modes}

Using the conformal relation \eqref{conf-g-ret} between the Schwarzschild and the conformal Schwarzschild GFs
together with their respective Hadamard forms \eqref{eqn:hadamardTGret} and \eqref{Gret-Had}, allows us to  write the direct part of the BPT GF as
\be
\Gd{s}(x,x'):= {}_sU(x,x')\delta_+(\sigma)=
\frac{1}{r\cdot r'}{_s{\hat{U}}}(x,x')\delta_+(\hat{\sigma}). \label{sGd-def} \ee
Then using Eqs.~\eqref{U-sep} and \eqref{uring-sol} we write the direct part of the BPT GF as 

\begin{eqnarray}
\Gd{s}(x,x')
&=&
\frac{1}{r\cdot r'}{_s{\bar{U}}}(x^i,x^{i'}){_s{\mathring{U}}}(x^A,x^{A'})\delta_+(\hat{\sigma})
\nonumber \\
&=& 
\frac{1}{r\cdot r'}{_s{\bar{U}}}(x^i,x^{i'})
e^{-is(\bar{\alpha}+\bar{\beta})}\left(\frac{\gamma}{\sin\gamma}\right)^{1/2}
\delta_+(\hat{\sigma}).
\label{eq:Had-F-GTd}
\end{eqnarray}

In general, ${}_s\bar{U}$ is given by \eqref{lambda-def}, and the expressions (\ref{lambda-final}) for $\lambda_s$, \eqref{vv-sol} for $\bar{\Delta}$ and \eqref{eq:bar-U} for $\bar{U}$ hold in the case where $r=r'$. See \cite{nolan2026spray} for the general case.

We define new, real-valued angular functions 
\be\label{eq:F-SWSH}
F_{\ell,s}\lP\theta\rP :=
\s Y_{\ell,-s}(\theta,\phi)e^{+is\phi}=
\s Y_{\ell,-s}(\theta,0),
\ee
which, clearly from Eq.~\eqref{eq:norm-SWSH}, obey the following orthonormality relations 
    \be\label{eq:norm-F}
\int_{-1}^{+1}\df\lP\cos\theta\rP
F_{\ell,s}\lP\theta\rP\,F_{\ell',s}\lP\theta\rP=2\pi \delta_{\ell\ell'}.
\ee
 We now equate Eqs.~\eqref{eq:Had-F-GTd} and \eqref{eq:mode-dec-GTd}, using \eqref{eq:F-SWSH}:
\be
{_s{\bar{U}}}(x^i,x^{i'})\left(\frac{\gamma}{\sin\gamma}\right)^{1/2}\delta_+(\hat{\sigma})
=   \sum\limits_{\ell=|s|}^{\infty}\sGTld{s}\lP r,r',\dt\rP\,\sqrt{\frac{2\ell+1}{4\pi}}
F_{\ell,s}\lP\gamma\rP
\ee   
We isolate for $\sGTld{s}$ by multiplying across by $F_{\ell',s}$, integrating over $\cos\gamma: -1\to +1$ and using \eqref{eq:norm-F} and \eqref{eq:F-SWSH}, to obtain:
\begin{eqnarray}\label{eq:Gld int}
  \sGTld{s}(r,r';\dt) &=& 2\pi
  \sqrt{\frac{4\pi}{2\ell+1}}
  {_s{\bar{U}}}(x^i,x^{i'})\theta(\dt)
  \int_{-1}^{+1} d(\cos\gamma) \delta(\hat\sigma)\left(\frac{\gamma}{\sin\gamma}\right)^{1/2}F_{\ell,s}\lP\gamma\rP
  \nonumber \\ 
  &=& 
2\pi
  \sqrt{\frac{4\pi}{2\ell+1}}
  {_s{\bar{U}}}(x^i,x^{i'})\left(\frac{\sin\tau}{\tau}\right)^{1/2}\theta(\dt)\theta(\pi-\tau) 
 \s Y_{\ell,-s}(\tau,0).
\end{eqnarray}

Note the use here of $\hat{\sigma}=-\tau^2/2+\gamma^2/2$ and $\tau>0$. Alternatively, \eqref{eq:Gld int} may be written in  terms of Jacobi polynomials 
and, for $m\geq |s|$, Lemma 13 of~\cite{michel2020mathematical})
\eqnalgn{\label{eq:SWSH-Jacobi}
    \sYlm(\theta,\phi)=N\lP\sin\lP\frac{\theta}{2}\rP\rP^{|m+s|}\lP\cos\lP\frac{\theta}{2}\rP\rP^{|m-s|} P_{\ell-\ell_0}^{(|m+s|,|m-s|)}(\cos\theta)e^{i m \phi},
}
where 
$$
    N:=(-1)^{\max(m,-s)}\sqrt{\frac{2\ell+1}{4\pi}}\sqrt{\frac{(\ell+\ell_0)!(\ell-\ell_0)!}{(\ell+\ell_1)!(\ell-\ell_1)!}},
$$
$\ell_0:=\max(|m|,|s|)$ and $\ell_1:=\min(|m|,|s|)$.
The normalization for the Jacobi polynomials is such that
\eqn{\label{eqn:normJacobiP}
    \int_{-1}^{1}(1-x)^\alpha(1+x)^\beta P_\ell^{(\alpha,\beta)}(x)P_{\ell'}^{(\alpha,\beta)}(x)\df x=\frac{2^{\alpha+\beta+1}}{2\ell+\alpha+\beta+1}\frac{\Gamma(\ell+\alpha+1)\Gamma(\ell+\beta+1)}{\ell!\,\Gamma(\ell+\alpha+\beta+1)}\delta_{\ell \ell'}.
}
Using Eq.~\eqref{eq:SWSH-Jacobi}, the expression \eqref{eq:Gld int} becomes 
\be\label{eq:Gsld expr}
 \sGTld{s}(r,r';\dt)=
2\pi\theta(\dt)\theta(\pi-\tau)
  {_s{\bar{U}}}(x^i,x^{i'})\left(\frac{\sin\tau}{\tau}\right)^{1/2} 
 \left(\cos\left(\frac{\tau}{2}\right)\right)^{|2s|}
 P^{(0,2|s|)}_{\ell-|s|}\lP\cos{\tau}\rP.
\ee
 Setting $s=0$, we see that \eqref{eq:Gsld expr} reduces to Eq.(20) of~\cite{CNOW2019}, up to an overall factor of $4\pi$ - see Footnote 3 above.

It is worth remarking on the two Heaviside $\theta$ distributions in Eqs.~\eqref{eq:Gld int} and \eqref{eq:Gsld expr}. The $\theta(\Dt)$ merely comes already from the Hadamard form \eqref{eqn:hadamardTGret}, serving to exclude the causal {\it future} -rather than past- of the field point. In its turn, $\theta(\pi-\tau)$ serves to ensure that $\sGTld{s}$ (and so also $\Gd{s}$), which lives in $\mathcal{M}_2$, is zero at times such that $\hat\sigma$ cannot be zero (and so they are outside the support of the direct term $\Gd{s}$) for any of the possible values of $\gamma\in [0,\pi]$ (see Fig.~3 in~\cite{casals2023global} for an illustration of this, which is in flat spacetime but nevertheless captures the main point). It is useful to consider the values that arise in a particular case. For the trajectory at fixed radius $r=6M$ considered above, we can solve $\tau=2\tau_1(\xi)=\pi$ numerically using the elliptic function representation (\ref{tau1-ell}). Then using \eqref{delta-t-elliptic}, we find that this corresponds to $\Delta t\simeq 20.72M$. 
As we see from Figure \ref{fig:error}, our small coordinate expansion (to $\mathcal(\Delta x)^{20}$) for $\bar\Delta$ maintains a high degree of precision (with relative error $\lesssim 10\%$) over the time regime where $\sGTld{s}$ has support, at least in the case of a worldline in Schwarzschild at constant $r=6M$.
We give a small coordinate expansion for  $\sGTld{s}$ in App.~\ref{sec:SmallCoordExpansionForBarU}. 

The modes  $\sGTld{s}$ may be calculated via Eqs.~\eqref{eq:Gld int} and \eqref{eq:Gsld expr} either exactly or approximately. In the exact calculation, ${_s{\bar{U}}}(x^i,x^{i'})$ is calculated using 
Eq.~\eqref{lambda-def}, with $\lambda_s$ and $\bar{U}$ therein, as well as $\tau$, calculated as per App.~\ref{sec:AppB}.
Alternatively, $\sGTld{s}$ may be calculated approximately using the high-order, small coordinate distance expansion in Appendix \ref{sec:SmallCoordExpansionForBarU}.

\begin{figure}
    \centering
    \subfloat{
        \includegraphics[width=0.48\linewidth]{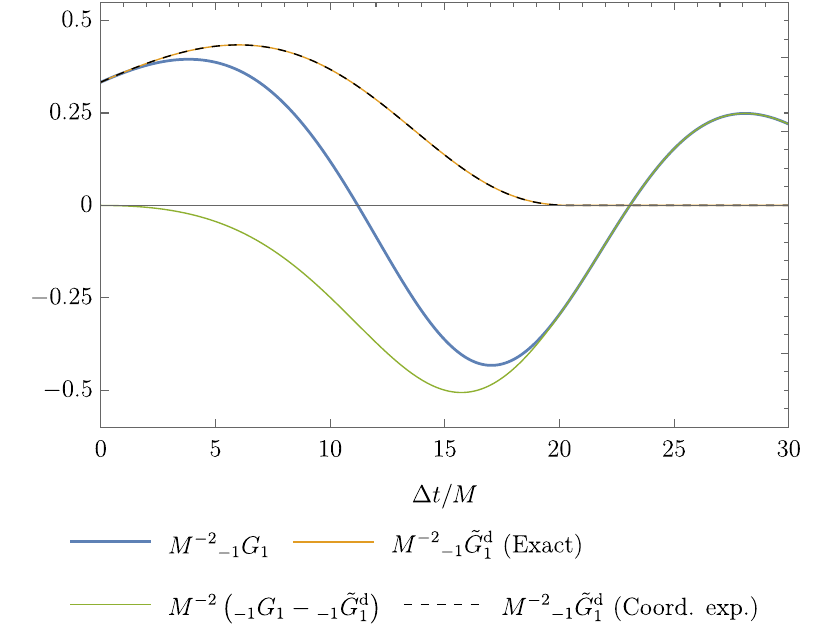}
    }
    \subfloat{
        \includegraphics[width=0.49\linewidth]{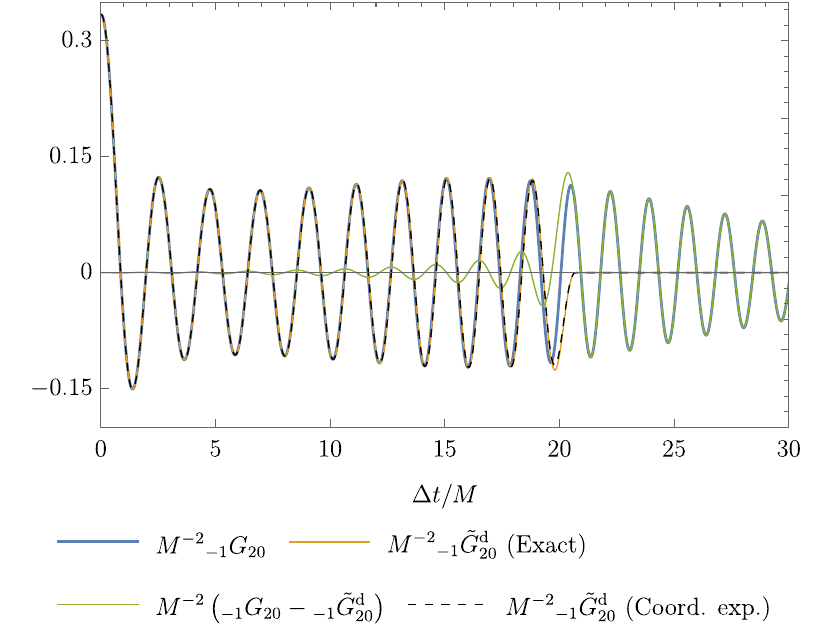}
    }
    \caption{Plots of the $\ell$-modes $\sGTl{-1}$ (blue), $\sGTldt{-1}$  (orange-solid for the exact and black-dashed for the expansion) and $\sGTl{-1}-\sGTldt{-1}$ (green)  of, respectively, the GF, its direct part and its non-direct part. They are plotted as functions of time  for $\ell=1$ (left)
    and $\ell=20$ (right)
    and $r=r'=6M$.}
    \label{fig:PlotBPTModeComparisonSpinMinus1}
\end{figure}

\begin{figure}
    \centering
    \subfloat{
        \includegraphics[width=0.48\textwidth]{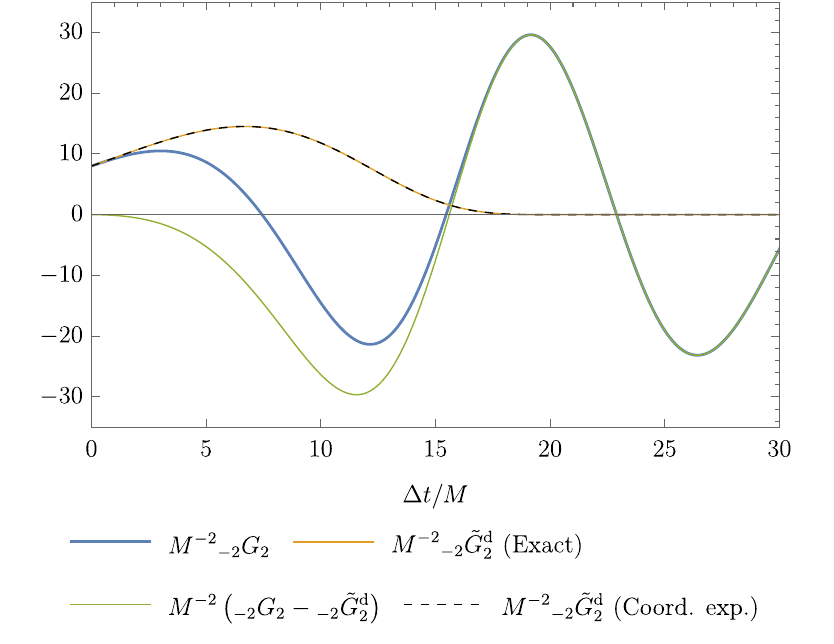}
    }
    \subfloat{
        \includegraphics[width=0.48\textwidth]{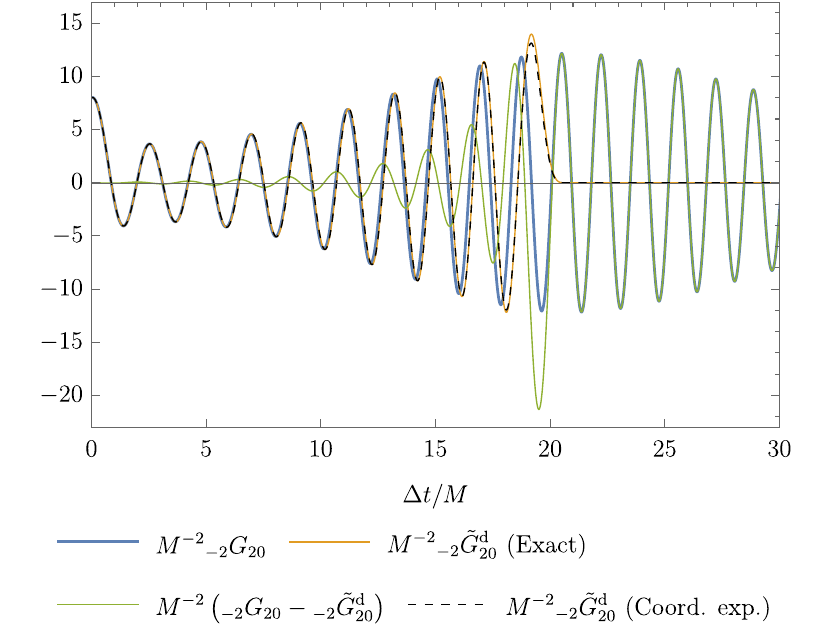}
    }
    \caption{Plots of the $\ell$-modes $\sGTl{-2}$ (blue), $\sGTldt{-2}$   (orange-solid for the exact and black-dashed for the expansion) and $\sGTl{-2}-\sGTldt{-2}$ (green)  of, respectively, the GF, its direct part and its non-direct part. They are plotted as functions of time  for $\ell=2$ (left) and $\ell=20$ (right)
    and $r=r'=6M$.}
    \label{fig:PlotBPTModeComparison}
\end{figure}

In~\cite{CNOW2019}, two of the authors together with two other researchers calculated the $\ell$-modes of the direct and non-direct parts in the case of a scalar field ($s=0$).
Here we  calculated them in the electromagnetic ($s=-1$) and gravitational ($s=-2$) field  cases.
In Fig.~\ref{fig:PlotBPTModeComparisonSpinMinus1} we show plots of $\sGTl{-1}$, $\sGTldt{-1}$ and their difference as functions of time in the case of constant $r=r'=6M$ for $\ell=1$ and $20$.
Fig.~\ref{fig:PlotBPTModeComparison} is a  similar plot but in the case of $s=-2$ (instead of $s=-1$) and for $\ell=2$ instead of $\ell=1$. 
Figs.~\ref{fig:PlotBPTModeComparisonSpinMinus1} and \ref{fig:PlotBPTModeComparison} plot $\sGTld{s}$ calculated both exactly (numerically) and approximately (via the small coordinate expansion to $\mathcal{O}(\Delta x^{20})$): there is visual agreement between the two for all times where these modes have support.

Figs.~\ref{fig:PlotBPTModeComparisonSpinMinus1} and \ref{fig:PlotBPTModeComparison} can be compared with the case $s=0$ in Fig.~1(b) in~\cite{CNOW2019} and with the spin-2 for the Regge-Wheeler equation (instead of BPT) in Fig.~3 in~\cite{aruquipa2026greenfunctionsreggewheelerteukolsky}, which has the same direct part as for spin-0.
One can see that, the larger the value of $\ell$, the larger the region of overlap between $\sGTl{s}$ and $\sGTldt{s}$.
This is unsurprising given that, as noted earlier, the $\delta(\sigma)$ divergence manifest in  Eq.~\eqref{eqn:hadamardTGret} for the GF $\s G_\textrm{ret}$, which is fully contained in its direct term $\Gd{s}$, arises from Eq.~\eqref{eq:mode-dec-GT} as a divergence in
the $\ell$-sum (when considered as a classical function). 

Eq.~\eqref{eq:Gsld expr} shows that   $\sGTld{s}$ are not smooth at $\tau=\pi$ due to the presence of the factor $\theta(\pi-\tau)$. However, they are continuous for all $s$ (thanks to the factor with $\sin\tau$) and the larger the value of $|s|$, the more regular they are at $\tau=\pi$.
The latter comment is thanks to the $\left(\cos(\tau/2)\right)^{2|s|}$, to the Jacobi polynomials not having a zero at $\tau=\pi$
and to ${}_s\bar{U}$ being smooth and nonzero at $\tau=\pi$ (as seen from the plots of $\tau$, $\bar{\Delta}$ and $\lambda$ in Figs.~\ref{fig:vv6M},  \ref{fig:sU} and \ref{fig:tau-lam}).
That the regularity of $\sGTld{s}$ increases with $|s|$ is manifest in Figs.~\ref{fig:PlotBPTModeComparisonSpinMinus1} and \ref{fig:PlotBPTModeComparison} for $s=-1,-2$ and in Fig.~1(b) in \cite{CNOW2019} for $s=0$ (even if the values of $\ell$ are also different).

\subsection{Calculation of the non-direct GF}

In the previous subsection we have calculated the $\ell$-modes $\sGTld{-2}$ of the direct part of the BPT GF.
In~\cite{aruquipa2026greenfunctionsreggewheelerteukolsky}, two of the authors calculated the BPT GF $\Gret{-2}$ and its $\ell$-modes $\sGTl{-2}$; we refer the reader to that paper for details of those calculations. 
Equipped with these, we are finally in a position to calculate the non-direct part $\Gnd{-2}$ of the GF using Eq.~\eqref{eq:decomp-nondirect}.

In Figs.~\ref{fig:LogPlotBPTGFComparison,circ} and \ref{fig:LogPlotBPTGFComparison,static} we show the comparison between $\Gret{-2}$, $\Gnd{-2}$ and $\Gd{-2}$.
We do so in the case that the spacetime points are, respectively, along: (i) a timelike circular geodesic at $r=6M$; and 
(ii) a static worldline at $r=6M$.
The red dots correspond to the value of ${}_{-2}V$ at coincidence (and so to the value of the GF as $\Dt \to 0^+$), given in Eq.~\eqref{eq:Vs-coinc}.
The plots of $\Gret{-2}$ were already presented in Figs.~15 and 16 in~\cite{aruquipa2026greenfunctionsreggewheelerteukolsky};
the plots of $\Gnd{-2}$ and $\Gd{-2}$ are new.
We note that we use in the $\ell$-sums for $\Gret{-2}$, $\Gnd{-2}$ and  $\Gd{-2}$ the smoothing factor in Eq.~(55) of ~\cite{aruquipa2026greenfunctionsreggewheelerteukolsky} with the values of 13 and 15 for $\ell_{\rm cut}$ in the circular and static cases, respectively.

Figs.~\ref{fig:LogPlotBPTGFComparison,circ} and \ref{fig:LogPlotBPTGFComparison,static} show that subtracting the direct part from the GF pushes the region of validity of the calculation to closer to coincidence ($\Dt=0$): from  $\Dt\simeq 4M$ to  $\Dt\simeq 1.6M$ in the circular case, and from  $\Dt\simeq 2.9M$ to  $\Dt\simeq 1M$ in the static case. The region of validity still does not  reach exactly coincidence, however.
We note that, even in the spin-0 case plotted in Fig.~2 in~\cite{CNOW2019}, the non-direct part also did not perform well for $\Dt \lessapprox 6M$, as noted in Sec.~IV.A  in~\cite{CNOW2019}. As explained there, the reason is that the non-direct part near coincidence is not equal to the Hadamard tail ${}_sV$ but to it times a $\theta_+$ distribution (see Eq.~\eqref{eq:Gnd Had}). Since in practise we can only compute up to a finite number of modes, only a smeared version of $\theta_+$ is in fact being computed, thus `contaminating' the calculation of the non-direct part at very early times.

Admittedly, despite the improvement we have achieved in representing the true GF for $\Dt>0$ from calculating the GF via the $\ell$-mode in sum~\eqref{eq:mode-dec-GT}  to calculating it as the non-direct part  via  the $\ell$-mode sum in~\eqref{eq:decomp-nondirect}, the improvement may not seem too spectacular at first sight. One should keep in mind, though, that, quantities like, for example, the self-force/field calculated as worldline integrals of the GF can be quite sensitive to small errors in the GF at small times. Indeed, practical calculations like  those in~\cite{CDOW13} show that the self-force/field are quite insensitive to the values of the GF near its singular support (where the $\ell$-sum in Eq.~\eqref{eq:decomp-nondirect} diverges) but sensitive to their values in-between, and also  insensitive to the value of the GF at late time (as it decays). Furthermore, since the non-direct part is not a good representation for the GF very near coincidence (as seen in Figs.~\ref{fig:LogPlotBPTGFComparison,circ} and \ref{fig:LogPlotBPTGFComparison,static} for $s=-2$ and in Fig.~2 in~\cite{CNOW2019} for $s=0$), in order to determine its value arbitrarily close to coincidence, one should probably calculate the GF very near coincidence in a rather different manner. This typically (see, e.g.,~\cite{CDOW13}) consists of calculating the Hadamard tail ${}_sV$ using a small coordinate expansion (similarly to what is done in App.~\ref{sec:SmallCoordExpansionForBarU} for ${}_s\bar U$) or a covariant  expansion (i.e., in terms of $\sigma^{\mu}$, similarly to what is done in App.~\ref{app:sVAtCoincidence} for ${}_sU$). Such calculation of ${}_sV$ very near coincidence would then be `matched' to the calculation of  the non-direct part achieved here. Currently, however, although a small coordinate expansion of ${}_sV$ for $s=0$ was achieved in~\cite{CDOWb}, no such expansion is yet known for $s\neq 0$. Also, a covariant expansion of ${}_sV$, such as the one in Eq.~(92)~\cite{2023PhRvD.108l5017I}, still requires the non-trivial calculation   of $\sigma^{\mu}$. Furthermore, achieving higher orders in a small coordinate expansion of ${}_sV$  (even for $s=0$) can be computationally quite expensive, and even more so in the case of covariant expansions. This is why obtaining a representation of the GF that is valid as close as possible to coincidence -- as we have done here -- is particularly useful.

\begin{figure}
    \centering
        \includegraphics[width=0.6\linewidth]{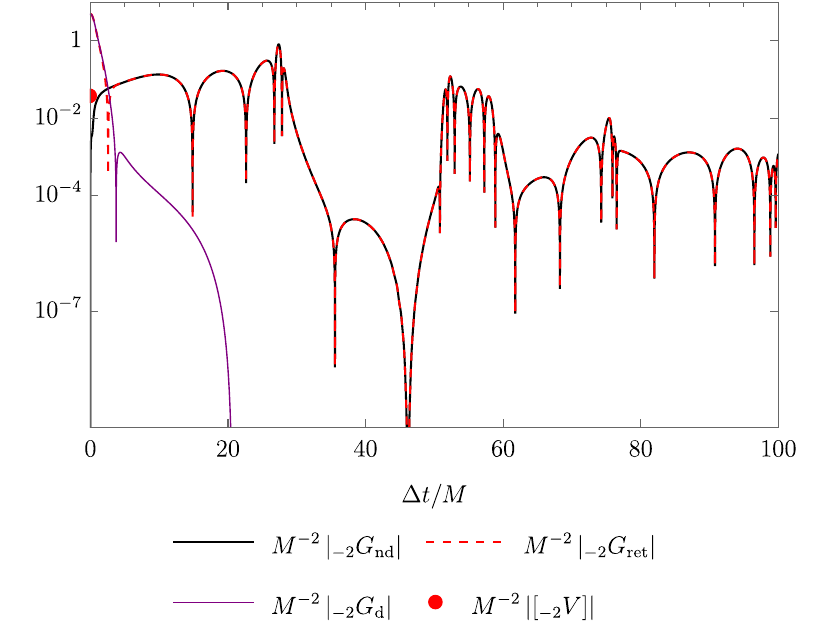}
    \caption{Log-plot of the BPT GF $\Gret{-2}$ (dashed red), its non-direct part $\Gnd{-2}$ (solid black) and its direct part  $\Gd{-2}$ (solid purple), along a timelike circular geodesic at $r=6M$. The red dot on the vertical axis is $[{}_{-2} V]$.}
    \label{fig:LogPlotBPTGFComparison,circ}
\end{figure}

\begin{figure}
    \centering
    \includegraphics[width=0.6\linewidth]{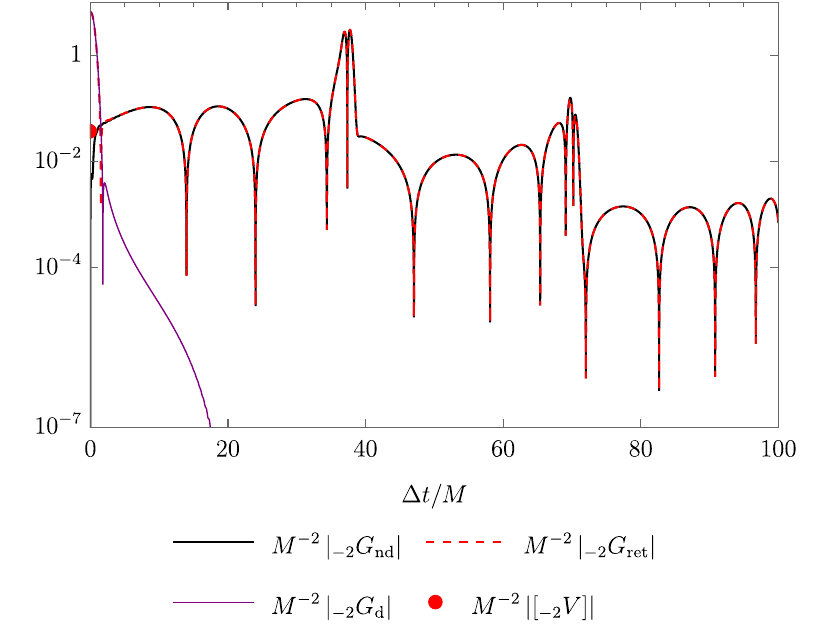}
    \caption{Log-plot of the BPT GF $\Gret{-2}$ (dashed red), its non-direct part $\Gnd{-2}$ (solid black) and its direct part  $\Gd{-2}$ (solid purple), along a static worldline at $r=6M$. The red dot on the vertical axis is the value of $[{}_{-2} V]$.}
    \label{fig:LogPlotBPTGFComparison,static}
\end{figure}

\section{Conclusions}

In this paper we have first analyzed the direct part of the Hadamard form for the GF of the Bardeen-Press-Teukolsky equation in Schwarzschild spacetime.
We started by obtaining closed-form expressions for this direct part for points along some specific worldlines (see Eqs.~\eqref{eq:Us circ} and \eqref{eq:UT,radial null}).

For {\it generic} spacetime points and in order to obtain the $\ell$-modes, we instead exploited the fact that Schwarzschild is conformal to $\hat{M}=\mathcal{M}_2\times \mathbb{S}^2$. In particular,  we have obtained an analytic and geometrical expression (Eq.~\eqref{eq:Had-F-GTd}) for this direct part in terms of the world function $\hat{\sigma}$ in $\hat{M}$,
a factor admitting a closed-form expression in terms of the Euler angles (see Eq.~\eqref{uring-sol}), the van Vleck determinant of $\mathcal{M}_2$ and and a factor  admitting an integral representation along geodesics in $\mathcal{M}_2$. In the case of constant radius worldlines of Schwarzschild, explicit formulas for these terms are given - see Eqs.~\eqref{lambda-def}, \eqref{eq:bar-U} and Eq.~\eqref{lambda-final}. These exact expressions - which can be generalised to essentially all pairs of points of Schwarzschild spacetime - are obtained by using the fact that $\mathcal{M}_2$ is a causal domain, and by introducing coordinates on $\mathcal{M}_2$ comprising proper time along the geodesics and the initial radial velocity \cite{nolan2026spray}.
We then used this expression for the direct part in order to obtain an expression (Eq.~\eqref{eq:Gsld expr}) for its multipolar $\ell$-modes and explicitly calculated them in the electromagnetic $s=-1$  (Fig.~\ref{fig:PlotBPTModeComparisonSpinMinus1}) and gravitational $s=-2$  (see Fig.~\ref{fig:PlotBPTModeComparison}) cases. 
We have derived (App.~\ref{sec:SmallCoordExpansionForBarU}) small coordinate distance expansions for the $\ell$-modes of the direct part and various other quantities in $\mathcal{M}_2$.

The $\ell$-modes of the direct part of the GF may  be useful in themselves for, e.g.,
modelling the early stage of the ringdown of a gravitational waveform (e.g.,~\cite{oshita2025probingdirectwavesblack,2026arXiv260122015S}).
In our case, we used these $\ell$-modes in order to calculate the non-direct part of the GF of the BPT equation for $s=-2$ (see Figs.~\ref{fig:LogPlotBPTGFComparison,circ} and \ref{fig:LogPlotBPTGFComparison,static}).
We have shown that this $\ell$-mode calculation of the non-direct part is a better representation than the corresponding $\ell$-mode calculation of the GF. 
The improvement, however and similarly to what already happened for $s=0$, does not reach all the way to coincidence. Thus, in order to calculate the GF very near coincidence, one would still need to supplement our calculation of the non-direct part with a separate calculation of the GF via, e.g., the Hadamard tail ${}_sV$ in Schwarzschild.  There are (at least) two routes available for such a calculation. First, note that thanks to our calculation of the non-direct part presented here, the calculation of  ${}_sV$ only needs to be achieved very near coincidence, so that a calculation of ${}_sV$ via some expansion (in terms of coordinate distances or the world function) would  only be necessary to a  lower order. 
Secondly, given that we have closed form expressions for the metric and for $\bar{U}$ on $\mathcal{M}_2$ in terms of the coordinates $(\tau,\xi)$, it may be possible to obtain results for ${}_sV$ through a combination of analytic and numerical approaches using those coordinates. We leave both such  calculations of ${}_sV$ for $s\neq 0$ for future work.

\section*{Acknowledgments}
BN acknowledges the hospitality of the Institut f\"ur Theoretische Physik, Universit\"at Leipzig and of the School of Theoretical Physics, Dublin Institute for Advanced Studies, where part of this work was completed.

\appendix
\section{Mollification of integrals in $\mathcal{M}_2$}\label{sec:AppA}

We collect some formulas relevant to the numerical evaluation of the integrals arising in Sec.~\ref{sec:direct-M2}. The quantities we wish to evaluate (and graph) are given in Eqs.~(\ref{tau-in-out}), (\ref{t-in-out}), (\ref{vv-sol}) and (\ref{lambda-final}). We have the following factorization of $R(\ep,r)$, which is defined in Eq.~\eqref{R-def}:
\be R=\ep r(r-r_{\rm{m}})(r-b)(r+r_{\rm{m}}+b)=:\ep(r-r_{\rm{m}})q^2(r,r_{\rm{m}}).\label{R-factor} \ee
Here $r_{\rm{m}}$ is the greater of the two positive roots of $R$ (which corresponds to the minimum of $r$ on the ingoing branch of the $\mathcal{M}_2$ timelike geodesics from $r_0>3M$) and we have 
\be b = \frac{r_{\rm{m}}}{2}\left(-1 + \left(\frac{r_{\rm{m}}+6M}{r_{\rm{m}}-2M}\right)^{1/2}\right).\label{b1-def} \ee

At the minimum radius $r=r_{\rm{m}}$, we have $\dot{r}^2=R=0$ and so we have the relations 
\be \ep^2 = \frac{f(r_{\rm{m}})}{r_{\rm{m}}^2},\quad \xi^2=\frac{r_0^2}{r_{\rm{m}}^2}(r_0^2f(r_{\rm{m}})-r_{\rm{m}}^2f(r_0)), \label{ep-xi-r1} \ee
and so (recalling that $\xi<0$ on the relevant geodesics) we can write
\be \xi = \xi(r_{\rm{m}}) = -\frac{r_0}{r_{\rm{m}}}(r_0-r_{\rm{m}})^{1/2} \left(r_0+r_{\rm{m}}-\frac{2M}{r_0r_{\rm{m}}}(r_0^2+r_0r_{\rm{m}}+r_{\rm{m}}^2)\right)^{1/2}.\label{xi-form} \ee 

The integrals $\tau_1(\xi)$, $t_1(\xi)$ and $\lambda=\lambda(\xi)$ are of the form (\ref{integrate-in-u}) and we can write 
\begin{eqnarray} \tau_1(\xi) &=& \frac{1}{\ep}\int_0^{(r_0-r_{\rm{m}})^{1/2}} Q(u,r_{\rm{m}})\,du,\label{tau1-int-u}\\
t_1(\xi) &=& \int_0^{(r_0-r_{\rm{m}})^{1/2}} \frac{(r_{\rm{m}}+u^2)^2}{f(r_{\rm{m}}+u^2)}Q(u,r_{\rm{m}})\, du, \label{t1-int-u} \\
\lambda(\xi) &=& -\int_0^{(r_0-r_{\rm{m}})^{1/2}} \frac{r_{\rm{m}}+u^2}{f(r_{\rm{m}}+u^2)}\left(1-\frac{3M}{r_{\rm{m}}+u^2}\right)Q(u,r_{\rm{m}})\, du, \label{lambda-int-u} 
\end{eqnarray}
where 
\be Q(u,r_{\rm{m}}) = \frac{2}{q(r_{\rm{m}}+u^2,r_{\rm{m}})}.\label{Q-q-def} \ee
From (\ref{ep-xi-r1}) we obtain
\be \frac{dr_{\rm{m}}}{d\xi}=-\xi\frac{r_{\rm{m}}^3}{r_0^4}\left(1-\frac{3M}{r_{\rm{m}}}\right)^{-1} \label{dr1-dxi} \ee
and we can use this to calculate (via the chain rule) 
\begin{eqnarray} \tau_1'(\xi) &=& -\frac{\xi}{\ep^2 r_0^4}\tau_1(\xi) +\frac12\frac{\xi r_{\rm{m}}^3}{\ep r_0^4}\left(1-\frac{3M}{r_{\rm{m}}}\right)^{-1}\left.Q(u,r_{\rm{m}})\right|_{u=(r_0-r_{\rm{m}})^{1/2}}(r_0-r_{\rm{m}})^{-1/2} \nonumber \\
&& -\frac{\xi r_{\rm{m}}^3}{\ep r_0^4}\left(1-\frac{3M}{r_{\rm{m}}}\right)^{-1}\int_0^{(r_0-r_{\rm{m}})^{1/2}} \frac{\partial}{\partial r_{\rm{m}}}\left(Q(u,r_{\rm{m}})\right) du. \label{dtau1-dxi} 
\end{eqnarray} 
Note that the (new) integral that arises is of the form (\ref{integrate-in-u}), and that the expression as a whole is smooth in the limit $r_{\rm{m}}\to r_0$. 

\section{Elliptic function representation}\label{sec:AppB}

We show here how the key integrals of Sec.~\ref{sec:direct-M2} can be written in terms of elliptic integrals. We recall that these integrals are 
\begin{eqnarray}
\tau_1 &=& \int_{r_{\rm{m}}}^{r_0} \frac{dr}{R^{1/2}(\ep,r)},\label{app-tau1} \\
\Delta t &=& 2\ep\int_{r_{\rm{m}}}^{r_0}\frac{r^2}{fR^{1/2}(\ep,r)}dr,\label{app-delta-t} \\
\lambda &=& -\int_{r_{\rm{m}}}^{r_0} \frac{\ep r}{fR^{1/2}}\left(1-\frac{3M}{r}\right)dr. \label{app-lambda} 
\end{eqnarray} 
We define $x=r_{\rm{m}}, x_0=r_0$ and the integrals 
\begin{eqnarray}
    J_1(x) &=& \int_x^{x_0} \frac{dr}{P^{1/2}}, \label{j1-def} \\
    J_2(x) &=& \int_x^{x_0} \frac{r^3}{r-2M} \frac{dr}{P^{1/2}}, \label{j2-def} \\
    J_3(x) &=& \int_x^{x_0} \frac{r(r-3M)}{r-2M}\frac{dr}{P^{1/2}} \label{j3-def}
\end{eqnarray}
where 
\be P = r(r-x)(r-b)(r+x+b). \label{q-def} \ee
The roots of this quartic are, in order, 
\be -x-b<0<b<x,\quad b=b(x) = \frac{x}{2}\left(-1+\left(\frac{x+6M}{x-2M}\right)^{1/2}\right). \label{roots} 
\ee
The pre-factors in the integrands of $J_2$ and $J_3$ may be written as
\begin{eqnarray}
    \frac{r^3}{r-2M} &=& 4M^2+2Mr+r^2 +\frac{8M^3}{r-2M},\label{j2-factor} \\
    \frac{r(r-3M)}{r-2M} &=& -M + r-\frac{2M^2}{r-2M}.\label{j3-factor} 
\end{eqnarray} 
Define 
\begin{eqnarray} I_j &=& \int_{x}^{x_0} \frac{r^j}{P^{1/2}}dr,\quad j=0,1,2, \label{Ij-def} \\
I_3 &=& \int_x^{x_0} \frac{dr}{(r-2M)P^{1/2}}. \label{I3-def} 
\end{eqnarray}
Then 
\begin{eqnarray} 
J_1 &=& I_0,\label{J1-I} \\
J_2 &=& 4M^2I_0+2MI_1+I_2+8M^3I_3,\label{J2-I} \\
J_3 &=& -MI_0+I_1-2M^2I_3,\label{J3-I} 
\end{eqnarray} 
and 
\be \tau_1 = \frac{1}{\ep}J_1,\quad \Delta t = 2J_2, \quad \lambda = -J_3. \label{key-in-J} \ee

To obtain the elliptic integral representation, we define $\psi\in[0,\pi/2]$ by 
\be \sin^2\psi = \left(\frac{x+2b}{2x+b}\right)\left(\frac{r-x}{r-b}\right) \in [0,1]. \label{psi-def} \ee
Equivalently, 
\be r = x\left(\frac{1-p\sin^2\psi}{1-n\sin^2\psi}\right) = \left(b+\frac{x-b}{1-n\sin^2\psi}\right),\label{r-psi} \ee
where
\be p = \frac{b(2x+b)}{x(x+2b)},\quad n = \frac{2x+b}{x+2b}.\label{d-n-def} \ee
Then it is straightforward to show that 
\begin{eqnarray} 
I_0 &=& \alpha F(\psi_0,p), \label{I0-ell} \\
I_1 &=& \alpha b F(\psi_0,p) + \alpha(x-b)\Pi(n;\psi_0,p) \label{I1-ell} 
\end{eqnarray} 
where $\psi_0=\left.\psi\right|_{r=r_0}$,
\be \alpha = \frac{2}{(x(x+2b))^{1/2}} \label{alpha-def} \ee
and $F$ and $\Pi$ are elliptic integrals of the first and third kind respectively: 
\begin{eqnarray}
    F(\psi_0,p) &=& \int_0^{\psi_0} \frac{d\psi}{\Delta_p^{1/2}},\label{EllipticF-def} \\
    \Pi(n;\psi_0,p) &=& \int_0^{\psi_0} \frac{d\psi}{\Delta_n\Delta_p^{1/2}}\label{EllipticPi-def} 
\end{eqnarray}
and where 
\be \Delta_z(\psi) = 1-z\sin^2\psi \label{delta-z-def} \ee
and the argument is taken to be $\psi$ when not specified. The substitution (\ref{r-psi}) in (\ref{Ij-def}) with $j=2$ shows that $I_2$ includes terms in $F(\psi_0,p)$, $\Pi(n;\psi_0,p)$ and in the integral 
\be \int_0^{\psi_0}\frac{d\psi}{\Delta_n^2\Delta_p^{1/2}}. \label{dn2-psi} \ee
For this, we apply the reduction formula Equation (336.02) of \cite{byrd2013handbook}. This results in 
\begin{eqnarray} I_2 &=& \frac{\beta}{2n^2(n-1)}
\Bigg(
n(p-n)E(\psi_0,p)+(n^2-2np+(2n-1)p^2)F(\psi_0,p)
\nonumber \\
&& \left. + (n-p)(n^2+2np-2n-p)\Pi(n;\psi_0,p) -n^2(p-n)\frac{\sin\psi_0\cos\psi_0\Delta_p^{1/2}(\psi_0)}{\Delta_n(\psi_0)}\right) \label{I2-ellip} 
\end{eqnarray}
where
\be \beta = 2\left(\frac{x^3}{x+2b}\right)^{1/2}\label{beta-def} 
\ee
and $E$ is the elliptic function of the second kind 
\be E(\psi_0,p) = \int_0^{\psi_0}\Delta_p^{1/2} d\psi. \label{EllipticE-def} 
\ee
To write $I_3$ in terms of elliptic functions, we note that 
\be \frac{1}{r-2M} = \frac{1}{b-2M}\left(1-\left(\frac{x-b}{x-2M}\right)\frac{1}{\Delta_m}\right), 
\label{I3-coefficient} \ee
where 
\be m = \frac{(2x+b)(b-2M)}{(x+2b)(x-2M)}. \label{mchar-def} \ee
This yields 
\be I_3 = \frac{\alpha}{b-2M}\left(F(\psi_0,p)-\left(\frac{x-b}{x-2M}\right)\Pi(m;\psi_0,p)\right). 
\label{I3-elliptic} \ee
Collecting the relevant terms, we have
\begin{eqnarray}
    \tau_1 &=& \frac{\alpha}{\ep}F(\psi_0,p), \label{tau1-ell} \\
    \Delta t &=& \frac{\beta(p-n)}{n(n-1)}E(\psi_0,p) +\left(4M\alpha(2M+b)+\frac{16M^3\alpha}{b-2M}+\frac{\beta}{n^2(n-1)}(n^2-2np+(2n-1)p^2)\right)F(\psi_0,p)\nonumber \\
    && + \left(4M\alpha(x-b)+\frac{\beta(n-p)}{n^2(n-1)}(n^2+2np-2n-p)\right)\Pi(n;\psi_0,p) \nonumber \\
    &&-\frac{16M^3\alpha}{b-2M}\left(\frac{x-b}{x-2M}\right)\Pi(m;\psi_0,p) -\frac{\beta(p-n)}{n-1}\frac{\sin\psi_0\cos\psi_0\Delta_p^{1/2}(\psi_0)}{\Delta_n(\psi_0)},\label{delta-t-elliptic} \\
    \lambda &=& -\alpha b\left(\frac{b-3M}{b-2M}\right)F(\psi_0,p) -\alpha(x-b)\Pi(n;\psi_0,p)-\frac{2M^2\alpha}{b-2M}\left(\frac{x-b}{x-2M}\right)\Pi(m;\psi_0,p).\label{lambda-elliptic} 
\end{eqnarray}
To calculate the van Vleck determinant in $\mathcal{M}_2$ in these terms, we recall (\ref{vv-sol}) which can be written as 
\be \bar\Delta = \frac{\tau_1(\xi)}{\xi\tau_1'(\xi)}.\label{vv-app} \ee
The numerator here is given by (\ref{tau1-ell}). To calculate the denominator, we note that we can write $\tau_1(\xi)=T_1(x)$ with 
\be \xi^2 = \ep^2r_0^4-r_0^2f(r_0),\quad \ep^2 =\frac{f(x)}{x^2}.\label{xi-x} \ee
We have the following partial derivatives of the elliptic function $F(\psi,p)$: 
\begin{eqnarray} 
\partial_\psi(F(\psi,p))&=& \frac{1}{\Delta_p^{1/2}},\label{d-psi-F} \\
\partial_p(F(\psi,p)) &=& \frac{1}{2p(1-p)}\left(E(\psi,p)-(1-p)F(\psi,p)-p\frac{\sin\psi\cos\psi}{\Delta_p^{1/2}}\right). \label{d-p-F}
\end{eqnarray}
Then we have 
\be \xi\tau_1'(\xi) = -\frac{\xi^2x^3}{r_0^4}\left(1-\frac{3M}{r}\right)^{-1}T'(x),\label{d-xi-tau1}
\ee
where
\begin{eqnarray} 
T'(x) &=&\tilde{\alpha}\left(\left(\frac{\tilde{\alpha}'}{\tilde{\alpha}}-\frac12\frac{p'}{p}\right)F(\psi_0,p)+\frac{p'}{2p(1-p)}E(\psi_0,p)+\frac{1}{\Delta_p^{1/2}}\left(\psi_0'-\frac{p'}{2(1-p)}\cos\psi_0\sin\psi_0\right)\right), \label{T-prime} 
\end{eqnarray}
\be \tilde{\alpha} = \frac{\alpha}{E} \label{tilde-alpha-def} \ee
and the primes in (\ref{T-prime}) are derivatives with respect to $x$.

To recap: (\ref{tau1-ell})-(\ref{vv-app}), together with (\ref{ep-xi-r1}), \eqref{d-xi-tau1}  and \eqref{T-prime}, provide elliptic integral representations of the proper time $\tau=2\tau_1$, the coordinate time $\Delta t$, the BPT parameter $\lambda$ and the van Vleck determinant $\bar{\Delta}$ in $\mathcal{M}_2$. These evaluate the relevant 2-point functions in the case where $r$ has the same value $r_0$ at the two Schwarzschild points $x^\alpha$ and $x^{\alpha'}$. The integrals depend on the fixed parameter $M$ and the constant radius $r_0$, and the running parameter $r_{\rm{m}}$, the minimum of $r$ on the unique $\mathcal{M}_2$ geodesic that connects the projection onto $\mathcal{M}_2$ of $x^\alpha$ to the corresponding projection of $x^{\alpha'}$. (Equivalently, we can consider $\xi$ to be the running parameter in the integrals.) The formulas apply in the case $r_0>3M$. 

\section{Small coordinate and near-coincidence expansions in $\mathcal{M}_2$}\label{sec:SmallCoordExpansionForBarU}

We start by expanding Synge's world function as
\eqn{
    \bar\sigma=\sum\limits_{i=0,k=0}^\infty{c}_{i k}(r)\Delta t^{i}\Delta r^k,
}
where $\Delta t=t-t'$, $\Delta r=r-r'$. The coefficients $c_{ik}$ are determined by inserting the above expansion into Eq.~\eqref{eq:bar-sigma} and we find
\eqnalgn{\notag
    \bar\sigma=\,&-\frac{f(r)}{2r^2}\Delta t^2+\frac{1}{2r^2f(r)}\Delta r^2-\frac{r-3M}{2r^4}\Delta t^2\Delta r+\frac{r-M}{2r^4f(r)^2}\Delta r^3-\frac{(r-3M)^2f(r)}{24r^6}\Delta t^4\\ \notag
    &-\frac{33 M^2-28 M r+5 r^2}{12 r^6 f(r)}\Delta t^2\Delta r^2+\frac{15M^2+11r^2f(r)}{24r^6f(r)^3}\Delta r^4-\frac{(r-3 M) \left(21 M^2-14 M r+2 r^2\right)}{24 r^8}\Delta t^4\Delta r\\\notag
    &+\frac{45 M^3-67 M^2 r+31 M r^2-4 r^3}{12 r^8 f(r)^2}\Delta t^2\Delta r^3+\frac{(r-M) \left(21 M^2-20 M r+10 r^2\right)}{24 r^8 f(r)^4}\Delta r^5\\\notag
    &-\frac{f(r) (r-3 M)^2 \left(45 M^2-30 M r+4 r^2\right)}{720 r^{10}}\Delta t^6-\frac{7065 M^4-10050 M^3 r+5087 M^2 r^2-1072 M r^3+78 r^4}{720 r^{10} f(r)}\Delta t^4\Delta r^2\\\notag
    &-\frac{945 M^4-2160 M^3 r+1781 M^2 r^2-606 M r^3+64 r^4}{240 r^{10} f(r)^3}\Delta t^2\Delta r^4\\
    &+\frac{945 M^4-2310 M^3 r+2251 M^2 r^2-1096 M r^3+274 r^4}{720 r^{10} f(r)^5}\Delta r^6+\mathcal{O}\lP\Delta x^7\rP,
    \label{eq:BarSigma-CoordExp}
}
which agrees --to the corresponding order-- with Eq.~(A1) in Ref.~\cite{CNOW2019}.

We similarly expand the direct Hadamard bitensor as
\eqn{
    {}_s\bar{U}=\sum\limits_{i=0,k=0}^\infty u_{ik}(r)\Delta t^i\Delta r^k,
}
for some coefficients $u_{ik}(r)$.
The coincidence limit $[{}_s\bar{U}]=1$ implies that $u_{00}=1$. We insert these expansions for ${}_s\bar{U}$ and $\bar\sigma$ into Eq.~\eqref{ubar-transport} to find the $u_{ik}$ coefficients recursively (similarly to $c_{ik}$). The result is
\eqnalgn{\notag
    {}_s\bar{U}=\,&1-\frac{s(r-3M)}{r^2}\Delta t-\frac{s(r-M)}{r^2f(r)}\Delta r+\frac{rf(r) (r-6 M) + 6 (r-3 M)^2 s^2}{12r^4}\Delta t^2\\\notag
    &-\frac{rf(r) (r-6M) s - 2 (r-M) (r-3 M) s^2}{2r^4f(r)}\Delta t\Delta r\\
    &-\frac{rf(r) (r-6 M) + 6 (2 M^2 - 2 M r + r^2) s - 6 (r-M)^2 s^2}{12r^4f(r)^2}\Delta r^2+\mathcal{O}\lP\Delta x^3\rP
    \label{eq:sU-CoordExp}
}

In Ref.~\cite{M2BPTHadamardU} we provide a Mathematica notebook (strongly based on the notebook ``TwodLocalExpansions\_arxiv.nb" in~\cite{CNOW2019} in the scalar case\footnote{The original version of ``TwodLocalExpansions\_arxiv.nb" in~\cite{CNOW2019} (which is for $s=0$) contains some errors which are corrected in our notebook in Ref.~\cite{M2BPTHadamardU} (which is for general $s$).}) which calculates the small coordinate separation expansions \eqref{eq:BarSigma-CoordExp} and \eqref{eq:sU-CoordExp} for, respectively, the world function and the direct Hadamard coefficient ${}_s\bar{U}$ in $\mathcal{M}_2$.
The notebook can generate expansions to arbitrary order 
$\mathcal{O}(\Delta x^{k})$ ($k=20$ was used in the numerical evaluations); here we only give it to $k=3$ for space considerations.
The higher-order coefficients are extremely unwieldy.

We can use the integrals of Appendix \ref{sec:AppA} to cross-check the coefficients in Eqs.~\eqref{eq:BarSigma-CoordExp} and~\eqref{eq:sU-CoordExp}. The base point of the small coordinate expansion (i.e.\ the point with $\Delta t= \Delta r=0$) is the initial point $(t,r)=(0,r_0)$ of the $\mathcal{M}_2$ geodesics that underpin the calculations of Section \ref{sec:Had} and Appendix \ref{sec:AppA}. Along the relevant world line in Schwarzschild, $\Delta r\equiv 0$. Small values of $\Delta t \gtrsim 0$ along the world line in Schwarzschild correspond to points whose projections onto $\mathcal{M}_2$ are connected by geodesics with turning point at $r_{\rm{m}}\lesssim r_0$. Thus the small coordinate expansions such as in~\eqref{eq:BarSigma-CoordExp} and~\eqref{eq:sU-CoordExp} with $\Delta r=0$ correspond to what we will call the \textit{near-coincidence expansion} that arises by approximating the integrals of Appendix \ref{sec:AppA} in the limit $\delta:=(r_0-r_{\rm{m}})^{1/2}\to 0$. To obtain these near-coincidence expansions for $\tau, \Delta t, \lambda_s$ and $\bar\Delta$, we expand the integrands of (\ref{tau1-int-u})-(\ref{lambda-int-u}) and (\ref{dtau1-dxi}) in powers of $u$, integrate, and expand in powers of $\delta$. This process yields expressions in the form of polynomials in $\delta$ of a chosen order (we work to order $N=24$) with $\mathcal{O}(\delta^{25})$ remainders. Among these we have the relations 

\begin{eqnarray} \bar{\Delta} &=& \sum_{k=1}^{24} \Delta_k\delta ^k + \mathcal{O}(\delta^{25}), \label{delta-delta-exp} \\
\Delta t &=& \sum_{k=1}^{24} \Delta t_k\delta ^k + \mathcal{O}(\delta^{25}) \label{delta-t-delta-exp} \end{eqnarray}
for some coefficients $\Delta_k, \Delta t_k$ depending on $r_0$ and $M$. Figure \ref{fig:vv6M} includes a parametric plot, with parameter $\delta$, of (the polynomial part of) $\bar{\Delta}$ against (the polynomial part of) $\Delta t$ (dashed blue curve).

We can compare the near-coincidence expansions in \eqref{delta-delta-exp} and  \eqref{delta-t-delta-exp} with the small coordinate expansions resulting from (\ref{eq:BarSigma-CoordExp}) and (\ref{eq:sU-CoordExp}) as follows. We invert (\ref{delta-t-delta-exp}) order by order to obtain the coefficients in the expansion
\be \delta = \sum_{k=1}^{24} \delta_k(\Delta t)^k + \mathcal{O}(\Delta t)^{25}. \label{delta-delta-t-exp} \ee
 We then substitute this expansion into the expansions for  $\bar{\sigma}=-\tau^2/2$ and ${}_s\bar{U}=e^{2s\lambda}\bar{\Delta}^{1/2}$ to obtain series expansions in terms of powers of $\Delta t$. We have verified that the resulting coefficients match those of (\ref{eq:BarSigma-CoordExp})  and (\ref{eq:sU-CoordExp}) to $\mathcal{O}(\Delta t^6)$. 

Given the small coordinate expansion for ${}_{s}\bar{U}$ and $\bar{\sigma}$, it is straightforward to obtain a small coordinate expansion for ${}_s G_\ell^\textrm{d}$. We simply insert Eq.~\eqref{eq:BarSigma-CoordExp} (considering $\tau=\sqrt{-2\bar{\sigma}}$) and Eq.~\eqref{eq:sU-CoordExp} into Eq.~\eqref{eq:Gsld expr}. In this way, for $r=r'=6M$ and $\Dt>0$, we find
\eqnalgn{\notag
    &\frac{{}_s G_\ell^\textrm{d}}{4\pi}=\frac{1}{2}-\frac{s}{24}\frac{\Dt}{M}-\frac{24 (2 \ell+1) | s| +12 \ell (\ell+1)+27 s^2+4}{5184 }\frac{\Delta t^2}{M^2}+\frac{s \left(24 (2 \ell+1) | s| +12 \ell (\ell+1)+33 s^2+4\right)}{62208 }\frac{\Dt^3}{M^3}\\\notag
    &+\frac{24 (2 \ell+1) | s|  \left(12 \ell (\ell+1)+18 s^2+1\right)+36 (38 \ell (\ell+1)+9) s^2+4 (3 \ell
   (\ell+1) (6 \ell (\ell+1)+1)-7)+27 s^4}{26873856 }\frac{\Dt^4}{M^4}-\\\notag
   &\frac{s \left(120 (2 \ell+1) | s|  \left(12 \ell (\ell+1)+30 s^2+1\right)+60 (126 \ell (\ell+1)+31)
   s^2+60 \ell (\ell+1) (6 \ell (\ell+1)+1)+2187 s^4-188\right)}{1612431360}\frac{\Dt^5}{ M^5}\\\notag
   &-\frac{1}{2437996216320 }\left[168 (2 \ell+1) | s|  \left(820 \ell (\ell+1) s^2+20 \ell (\ell+1) (6 \ell (\ell+1)-5)-405 s^4+3 \left(5
   s^2-49\right)\right)\right.\\\notag
   &+252 (5 \ell (\ell+1) (134 \ell (\ell+1)-31)-198)
   s^2+84 \ell (\ell+1) (10 \ell (\ell+1) (4 \ell (\ell+1)-5)-147)\\
   &\left.+1260 (69 \ell (\ell+1)+10) s^4-79947 s^6-436\right]\frac{\Dt^6}{M^6}+\mathcal{O}\left(\frac{\Delta t^7}{M^7}\right).
   \label{eq:Gld-exp}
}
We have checked that ${}_s G_\ell^\textrm{d}/4\pi$ (the factor $4\pi$ is due to the different definition of direct modes in the two papers)  agrees with Eq.~(25) in~\cite{CNOW2019} when $s=0$ (and $M=1$).

\section{$\s V$ at coincidence}
\label{app:sVAtCoincidence}
In order to derive the value in Eq.~\eqref{eq:Vs-coinc} of the Hadamard tail coefficient $\s V(x,x')$ at coincidence, we first calculate a covariant expansion for $\s U$, which is valid for the operator in \eqref{sT-op} for generic $A_{\mu}$ and $B$. We write
\eqn{\label{eqn:sUCovExp}
    \s U(x,x')=u(x)+u_\mu(x)\sigma^\mu(x,x')+\frac{1}{2}u_{\mu\nu}(x)\sigma^\mu(x,x')\sigma^\nu(x,x')+\mathcal{O}(\eta^3),
}
where $\eta$ represents the size of a typical component of $\sigma^\mu$  and the coefficients $u(x)$, $u_\mu(x)$ and $u_{\mu\nu}(x)$ are calculated by inserting the above expansion into Eq.~\eqref{eqn:transportUT} and solving the resulting equation, order by order in $\eta$, and recursively for each coefficient. In this way we find:
\eqnalgn{
    u(x)=\,&1,\\
    u_\mu(x)=\,&-\frac{1}{2}A_\mu,\\
    u_{\mu\nu}(x)=\,&\frac{1}{2}\nabla_{(\mu}A_{\nu)}+\frac{1}{4}A_{\mu}A_{\nu}+\frac{1}{6}R_{\mu\nu},
}
and, consequently,
\eqn{\label{eq:sU-exp1}
    \s U=1-\frac{1}{2}A_\mu\sigma^\mu+\frac{1}{4}\lP\nabla_{(\mu}A_{\nu)}+\frac{1}{2}A_{\mu}A_{\nu}\rP\sigma^\mu\sigma^\nu+\frac{1}{12}R_{\mu\nu}\sigma^\mu\sigma^\nu+\mathcal{O}(\eta^3).
}
This expansion was already independently obtained in Ref.~\cite{2023PhRvD.108l5017I} with $R_{\mu\nu}=0$; our formula generalizes it to the case of non-zero Ricci tensor.
When inserting this expansion for $\s U$ in Eq.~\eqref{eqn:hatVTsInitialCondition}, we obtain
\eqn{\label{eq:[sV]}
    [\s V]=-\frac{1}{4}\nabla_\mu A^\mu-\frac{1}{8}A_\mu A^\mu+\frac{1}{2}B+\frac{R}{12}=\frac{R}{12}-s^2\frac{2M}{r^3}=-s^2\frac{2M}{r^3},
}
where in the last equality we use the fact that $R=0$ in Schwarzschild.
Eq.~\eqref{eq:[sV]} is  consistent with Eq. (92) in \cite{2023PhRvD.108l5017I}, with a different normalization factor $\alpha=-\frac{1}{8\pi^2}$, in the generic-$s$ case and it agrees with Ref.~\cite{poisson2011motion} in the $s=0$ case (where $[{}_0 V]=\frac{R}{12}=0$).


\bibliographystyle{apsrev}
\bibliography{mybib_BN}

\end{document}